\documentclass[useAMS,usenatbib]{mn2e}

\usepackage{psfig, epsf, epsfig}

\title[Formation of globular clusters]{Globular cluster
formation  with multiple
stellar populations from
 hierarchical star cluster complexes} 
\author[K. Bekki]
{Kenji Bekki${}^1$\thanks{E-mail:
kenji.bekki@uwa.edu.au} \\
${}^1$ICRAR M468
The University of Western Australia
35 Stirling Hwy, Crawley
Western Australia 6009, Australia}

\begin{document}

\date{Accepted, Received 2005 February 20; in original form }

\pagerange{\pageref{firstpage}--\pageref{lastpage}} \pubyear{2005}

\maketitle

\label{firstpage}

\begin{abstract}

Most old globular clusters (GCs) in the Galaxy are observed to have internal
chemical abundance spreads in light elements.
We discuss a  new GC formation scenario  based on hierarchical star formation
within fractal molecular clouds.
In the new scenario, 
a cluster of bound and unbound  star clusters (`star cluster complex', SCC)
that have a power-law cluster mass
function with a slope ($\beta$) of 2
is first formed from a massive gas clump
developed 
in a dwarf galaxy.
Such cluster complexes  and
$\beta=2$ are observed and expected from hierarchical star formation.
The most massive star cluster  (`main cluster'),
which is the progenitor of a GC,
can accrete gas ejected from asymptotic giant branch (AGB) stars 
initially in the cluster and 
other low-mass clusters before the clusters are tidally stripped
or destroyed to become field stars in the dwarf.
The SCC is initially embedded in a giant gas hole created by numerous supernovae of the SCC
so that cold gas outside the hole can be accreted onto the main cluster later.
New stars formed
from the accreted gas have 
chemical abundances that are different from those of the original SCC. 
Using hydrodynamical simulations of GC formation based on this scenario,
we show that the 
main cluster with the initial mass as large as $[2-5] \times 10^5 {\rm M}_{\odot}$ 
can accrete more than $10^5 {\rm M}_{\odot}$ gas from  AGB stars of the SCC.
We suggest that  merging of
hierarchical star cluster complexes can play key roles in
stellar halo formation around GCs and self-enrichment processes
in the early phase of GC formation.

\end{abstract}


\section{Introduction}

One of remarkable discoveries in the field of globular clusters (GCs) is that
old GCs in the Galaxy and intermediate-age ones in the Large Magellanic
Cloud (LMC) have multiple stellar populations
(e.g., Freeman \& Rodgers 1975; Cohen 1981;  Lee at al. 1999;
Gratton et al. 2001; Bedin et al. 2004; Norris 2004; Piotto et al. 2005; 
Mackey \& Broby Nielsen 2007; Lee et al. 2009;  Da Costa et al. 2014;
See Gratton et al. 2012 for a recent review). 
Most of the investigated GCs in the Galaxy are observed to show anti-correlations
between light elements (e.g., C, N, and O) of cluster members stars 
(e.g., Carretta et al. 2009; C09)
whereas only 8 GCs have been so far confirmed to have star-to-star abundance spreads
in heavy elements (e.g., Yong et al. 2014; Marino et al. 2015). 
Extended  main-sequence turn-offs (eMSTOs) and splits in main-sequence 
observed in the color magnitude diagrams (CMDs) of some LMC GCs
(e.g., Mackey \& Broby Nielsen  2007; Goudfrooij et al. 2014; Milone et al. 2016)
can be possible evidence for the multiple stellar populations
with different ages,
though recent observations suggest that internal stellar rotation
rather than age spreads
could explain the physical properties of LMC clusters with 
eMSTOs  (e.g., Bastian \& De Mink 2009;
Milone et al. 2016; Li et al. 2016).
These new discoveries stimulated much discussion on the initial stellar mass function
of stars in GCs, the formation processes of
GCs, and the origin of the observed diversity in chemical and dynamical properties
of GCs with multiple stellar populations
(e.g., D'Antona\&  Caloi 2004; Bekki et al. 2007;
Baumgardt et al. 2008;   D'Ercole et al. 2008, D08;
Vesperini et al. 2010; Renzini 2015;  D'Antona et al. 2016).

A straightforward scenario for this multiple stellar population
phenomenon in GCs is that gas ejected from the first generation (FG) of stars in
a GC is
converted into the second generation (SG) of stars with chemical abundances different from
those of FG stars: multiple populations mean multiple generations of stars.
In this scenario,  chemical abundances between the two generations
are different because chemical abundances of gas ejected from fast-rotating massive
stars (e.g.,Decressin et al. 2007),
supermassive stars (e.g., Denissenkov \& Hartwick 2014), 
massive interacting binaries (e.g., Bastian et al. 2013),
and AGB stars (e.g.,D08) are quite
different from the averaged ones of FG stars.
This scenario has been suggested to have a number of serious problems in explaining
the fundamental properties of GCs, for example, the larger fraction of SG stars in
GCs with multiple stellar populations. For this scenario to explain the observed
fraction of SG stars ($\sim 70$\%; C09),
the original mass of FG stars ($M_{\rm FG}$) should be much more massive
than the present-day GC mass (`mass budget problem').

This mass budget problem can be simply formulated as follows:
\begin{equation}
M_{\rm FG}=4.7 \times 10^6 
{ ( \frac{ \epsilon_{\rm sf} }{0.3} ) }^{-1}
{ ( \frac{ f_{\rm ej} }{0.1} ) }^{-1} 
( \frac{ M_{\rm SG,0} }{1.4 \times 10^5 {\rm M}_{\odot} } )
{\rm M}_{\odot},
\end{equation}
where $M_{\rm SG,0}$ is the present-day total mass of SG stars in a GC,
$\epsilon_{\rm sf}$ is the star formation efficiency in the SG star
formation ($M_{\rm SG}/M_{\rm ej}$, where $M_{\rm ej}$ is the total
mass of gas ejected from `polluter', such as AGB stars),
and $f_{\rm ej}$ is the mass fraction of gas ejected from polluters 
in the FG stars. In this estimation of $M_{\rm FG}$,
all SG stars are assumed to be long-lived low-mass stars that are alive in the present,
which means that a very unique initial mass function
(IMF) of stars  (i.e., bottom-heavy and top-light IMF) 
is assumed just for clarity.  Nevertheless, the required $M_{\rm FG}$ is much larger than
both the present-day mass of FG stars 
($M_{\rm FG,0}=6.0 \times 10^4 {\rm M}_{\odot}$ for a typical present-day GC mass
of $M_{\rm gc}=2 \times 10^5 {\rm M}_{\odot}$)
and the typical mass of GCs.
Although the mass budget problem is more complicated than the above discussion,
it is  one of the most serious problems in previous GC formation scenarios.
Smith \& Norris (1982) first discussed this mass budget problem in the context
of the origin of CN-weak and CN-strong populations in NGC 6752 and 47 Tuc,
though they did not use the term `mass budget problem'.

There are three scenarios  to solve the mass budget problem
(e.g., D'Antona \& Caloi 2004;
Bekki \& Norris 2006; Prantzos \& Charbonnel 2006).
First is that GCs were initially formed as very massive
star clusters composed only of FG stars
and then lost preferentially  the large fraction of FG   stars
by some physical processes (`VMSC' scenario).
Second is that GCs were  stellar galactic nuclei
of nucleated dwarfs that had been completely destroyed by their host galaxies'
tidal fields (`SGN' scenario).
Third is that the IMF of FG stars is top-heavy (i.e., a larger mass
fraction of massive stars and intermediate-mass stars
that eject gas) whereas that of SG stars
is bottom-heavy (i.e., a larger number of low-mass stars) for some physical reasons
(`top-heavy IMF' scenario).
Although GC formation processes  based on the VMSC scenario
have been extensively investigated by several authors
(e.g., D08; Bekki 2011, B11), recent observations have suggested potentially serious
problems of the VMSC scenario (e.g., Larsen et al. 2012).
Although some massive  GCs such as  $\omega$ Cen could have been formed from nucleated
dwarf galaxies (e.g., Freeman 1993; Bekki \& Freeman 2003), it is not clear
whether typical GCs can be formed in the SGN scenario.
The top-heavy IMF scenario has not clearly explained why
the IMFs of FG stars can be  top-heavy  in GC formation.

The Tarantula Nebula (a.k.a `30 Dor') 
in the LMC would provide a
hint for the solution of the mass budget problem as follows.
30 Dor with a diameter of $\sim 200$ pc contains a central main cluster `R136', 
other low-mass clusters such as Hodge 301 and NGC 2060,
numerous small stellar clumps and star-forming regions, and older pre-main sequence 
(PMS) stars
(e.g., Grebel \& Chu 2000; De Marchi et al. 2011  Sabbi et al. 2013).
De Marchi et al. (2011) analyzed the ages of PMS stars in 30 Dor 
and found that (i) there are
several generations  of stars with ages ranging from 1 to 30 Myr
and (ii) older PMSs are mostly located in the outer (eastern) part of R136.
These observations strongly suggest that a massive star cluster R136 can 
form with unbound stellar associations (that has become numerous field stars now)
and low-mass clusters around R136 
with a time scale of $\sim 30$ Myr. 
These furthermore imply that massive young star clusters,
which can be the progenitor of old GCs,
can form with other numerous unbound stellar association and
low-mass clusters within massive giant molecular clouds (GMCs).

Clustering of star clusters is observed in the star-forming regions
of the Galaxy and the LMC, 
nearby star-forming galaxies such as M33 and M51,
and  galaxy mergers
(e.g., Efremov 1995; Efremov \& Elmegreen 1998;
Zhang et al. 2001; Larsen 2004; Bastian et al. 2005;
Elias et at. 2009; Adamo et al. 2012).
These star cluster complexes (`SCCs') have been observationally investigated for
some specific cases, and their physical properties have been derived.
For example, 
Scheepmaker et al. (2009) investigated the two-point autocorrelation function
for three different groups of SCs with different age ranges
in M51 and found that 
that the projected fractal dimensions of $1.2-1.6$ can well describe
the observed slopes of the autocorrelation functions.
Bastian et al. (2005) revealed that SCCs in M51 are all younger than
10 Myr and have sizes of $85-240$pc and masses 
of $[3-30]\times 10^4 {\rm M}{\odot}$.  
The slopes of power-law cluster mass functions (i.e., $N_{\rm sc} \propto m_{\rm sc}^{-\beta}$,
where $m_{\rm sc}$ is a SC mass)
have been also investigated by many authors,
and $\beta$ appears to be approximately 2 for SCs in different galaxies
(e.g., Battinelli et al. 1994; Elmegreen \& Efremov 1997; de Grijs et al. 2003).
Such clustering of SCs with $\beta=2$ is a natural result of SC formation
from hierarchical star formation (See Elmegreen 2008 for a review).
It is therefore highly likely that the progenitor clusters
of the present-day GCs  were initially members of SCCs at their birth.

Such possibly more realistic GC-forming environments
within SCCs were not considered
in previous GC formation models 
(e.g., D08; B11) in which only AGB ejecta from GC progenitor massive single
clusters can be converted into SG stars (`self-accretion'`).
If gas ejected from  massive AGB stars in
unbound and bound low-mass clusters
surrounding  a very young massive  SC (`main cluster')
can be accreted onto the main cluster (`external accretion'), then
the total mass of these AGB ejecta can be a significant fraction of the total
mass of the main cluster.
Accordingly, the mass budget problem can be much less severe in the new
GC formation from SCCs.
However, 
low-mass clusters that form with the  main cluster 
can be quickly stripped from the surrounding of the main cluster
and subsequently destroyed by the tidal fields of their host dwarf galaxy
and even by the main cluster itself.
Therefore, it is possible that only a small fraction of their AGB ejecta
could be accreted onto the main cluster before they are stripped or disintegrated
to become field stars.
It is thus worthwhile 
to investigate how much  of gas ejected from AGB stars born
in low-mass clusters  can be  accreted
onto the main cluster  in the new SCC scenario. 

\begin{figure*}
\psfig{file=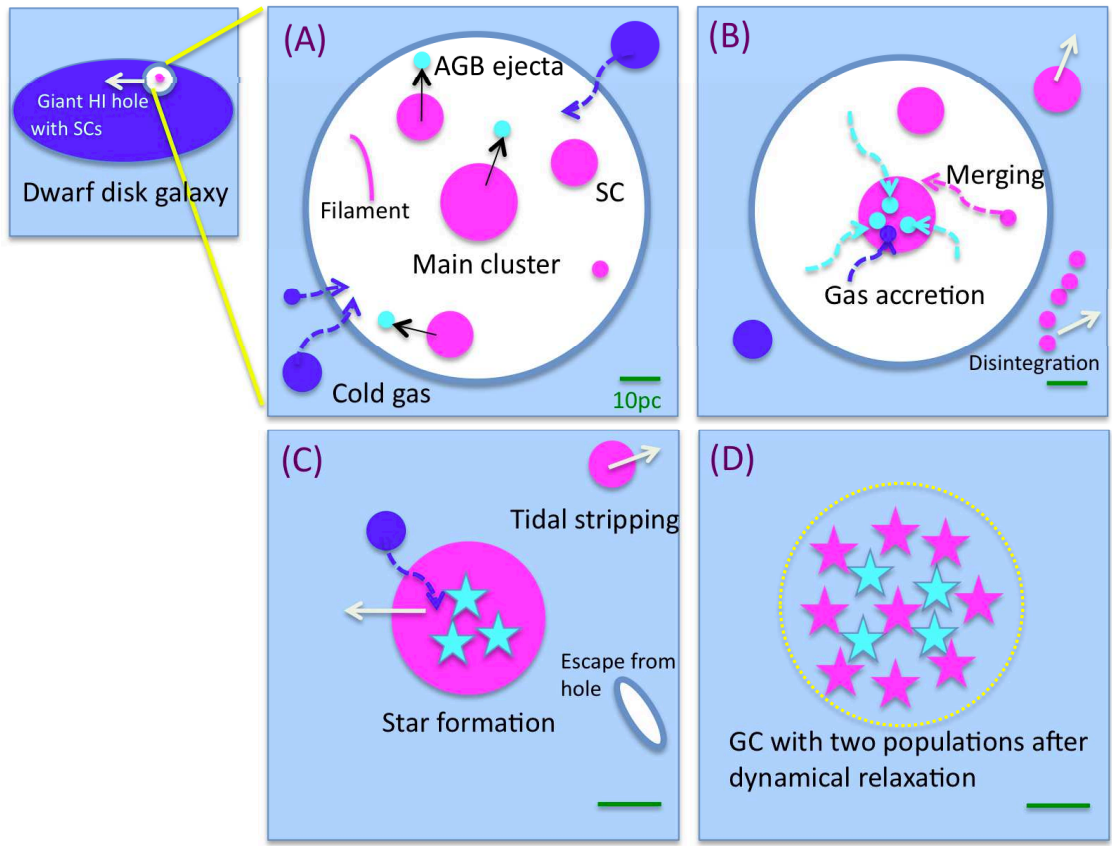,width=15.0cm}
\caption{
A brief illustration of the new scenario of GC formation from SCCs.
The selected key four phase of GC formation processes are shown in 
the four panels, {\bf (A)-(D)}, in chronological order. After the formation
of a SCC embedded in a giant gas hole,  AGB stars in the main cluster
and other low-mass clusters eject  gas to the
intra-cluster region through stellar winds {\bf (A)}. The AGB ejecta can be
accreted onto the main cluster very efficiently while the cluster is in
the gas hole {\bf (B)}.  Cold ISM can be accreted onto the main cluster well
after the cluster escapes from the hole (or the gas hole disappears through
gaseous dissipation) so that star formation can become efficient in
the inner region of the cluster {\bf (C)}. Most of the low-mass clusters
are either stripped from the surrounding of the main cluster or disrupted
by the tidal field of the host dwarf galaxy {\bf (D)}. Some low-mass clusters
merge with the main cluster to be destroyed by the cluster.
The remnants of the low-mass clusters become  a diffuse stellar halo around
the main cluster. The final cluster 
can appear as a GC that has  a diffuse stellar halo and two  stellar populations
with different chemical abundances. 
}
\label{Figure. 1}
\end{figure*}

The purpose of this paper and our forthcoming
papers  is  to investigate
the  formation of  GCs with multiple stellar populations 
from SCCs with $\beta=2$ that are expected from hierarchical
star formation within fractal molecular clouds.
Using hydrodynamical simulations of dwarf galaxies with SCCs,
we particularly investigate how much AGB ejecta from low-mass SCs that form
with the main clusters  
can be accreted onto the main clusters  within $\sim 300$ Myr.
This possible gas accretion timescale of $\sim 300$ Myr is chosen,
firstly because low-mass SCs in SCCs 
are expected to merge with one another or be stripped from
the main clusters within  $\sim 300$ Myr,
and secondly because recent observations of the LMC GCs
suggested that typical age spreads in the GCs are 100-400 Myr
(e.g., Goudrouij et al. 2014).
It should be noted here that it is still  controversial whether
the LMC clusters can contain  stellar populations with different ages
(e.g., Milone et al. 2016; Li et al. 2016).
If the total mass of AGB ejecta accreted onto main clusters  ($M_{\rm acc}$) is
significantly more than $10^5 {\rm M}_{\odot}$ (corresponding to the total mass
of SG stars in typical GCs),
then the mass budget problem
is much less serious
in the SCC scenario.
We investigate $M_{\rm acc}$ for different initial main cluster
masses ($m_{\rm mc}$) in SCCs,
initial positions of the SCCs within their host dwarfs,  and presence
or absence of cold interstellar medium (ISM) in dwarf galaxies.

The plan of the paper is as follows.
We outline the new scenario of GC formation from SCCs in dwarf galaxies
in \S 2.
We describe the models for dwarf galaxies, SCCs, and 
gas accretion onto main clusters 
in \S 3.
We present the key results of the simulation, in particular,
the time evolution of gas accretion rates and total masses of gas
accreted onto main clusters in \S 4.
We briefly discuss the important  implication of the present  results
in the context of the observed physical properties of GCs 
with multiple stellar populations in \S 5.
We summarize our  conclusions in \S 6.
It should be stressed here  that 
this paper is the very first step toward the better understanding
of GC formation in the context of the SCC scenario.
Therefore, we investigate only one of the most important physical
processes of GC formation in the present study.
We will discuss other importance processes of GC formation in
our forthcoming papers.

It is being hotly debated  whether the observed eMSTOs and splits of 
main-sequence of the LMC clusters can result from age spreads
or from stellar rotation (e.g., Milone et al. 2016; Li et al. 2016).
Recently,  For \& Bekki (2016) have 
discovered  direct evidence for ongoing star formation
(i.e., the presence of young stellar objects)  in  the older LMC clusters
with ages of $0.1-1$ Gyr. Their results strongly suggest that at least
some of the LMC clusters experienced secondary star formation after
the main initial burst of star formation.
They have also suggested that AGB ejecta needs to be accreted
onto the older clusters by some physical mechanisms. They also have found that
even some low-mass SCs with the masses less than $10^4 {\rm M}_{\odot}$
can have  ongoing star formation.
The present results will be able to provide a new clue to the origin of these
observations.
Our recent simulations have found that
(i) multiple stellar structures in FG stellar systems of
the simulated GCs can be formed from
massive gas clumps developed in gas-rich dwarf galaxies
and (ii) some of the SG stars of the GCs can be  formed
from gas that are not from AGB stars
of the  FG systems  but from the surrounding field AGB stars that simultaneously
form with the FG stars
(Bekki 2015a, 2016).
The present study is motivated by these recent results.

\begin{table}
\centering
\begin{minipage}{85mm}
\caption{Description of the basic parameter values
for the dwarf disk galaxy models.}
\begin{tabular}{llllll}
{ Model ID \footnote{ The dwarf model used in the fiducial model
is DW1. The models with $c=5.4$ and $r_{\rm s}=1$ kpc
correspond to high-z more compact
dwarf models.
 }} & 
{ $M_{\rm h}$ \footnote{ The initial total mass for dark matter halo
in units of $10^{10} M_{\odot}$.
 }} & 
{ $M_{\rm s}$  \footnote{  The initial total mass for stellar disk
in units of $10^{10} M_{\odot}$. The value of $M_{\rm s}$ in DW1 
is referred to as $M_{\rm s,0}$ for convenience.
}} &
{ $M_{\rm g}$  \footnote{ The initial total mass for gas disk
in units of $10^{8} M_{\odot}$.
}} &
{ $c$  \footnote{ The central concentration parameter in the 
in NFW dark matter profile.
}} &
{ $R_{\rm s}$  \footnote{ The initial size of stellar disk.
in units of kpc.
}}  \\
DW1 & 1.0 & 6.0 & 0.0 & 16.0 & 1.75\\
DW2 & 1.0 & 3.0 & 0.0 & 16.0 & 1.75\\
DW3 & 1.0 & 0.6 & 0.0 & 16.0 & 1.75\\
DW4 & 1.0 & 6.0 & 0.6 & 16.0 & 1.75\\
DW5 & 1.0 & 0.6 & 0.6 & 16.0 & 1.75\\
DW6 & 1.0 & 3.0 & 0.0 & 5.4 & 1.0 \\
DW7 & 1.0 & 3.0 & 0.6 & 5.4 & 1.0 \\
DW8 & 1.0 & 0.6 & 0.6 & 5.4 & 1.0 \\
\end{tabular}
\end{minipage}
\end{table}

\section{The new scenario}

The new SCC scenario is based both on the observed properties of
30 Dor and  SCCs in nearby galaxies (e.g., Efremov 1995;  Bastian et al. 2005;
Sabbi et al. 2013; Adamo et al. 2012)
and on our recent numerical simulations of GC formation
(Bekki 2015a; Bekki 2016).
In the new scenario,  a GC was initially the most massive star cluster (SC)
that formed with other numerous stellar associations and low-mass clusters
within a massive GMC (or a GMC association)
with fractal structures  in a gas-rich dwarf galaxy. 
Therefore, the formation processes of GCs in the SCC scenario, such as gas accretion
onto existing stellar systems and the subsequent (secondary) star formation within them,
could be significantly different from those described in previous models
(e.g., D08; B11).
The new scenario consists of the following seven stages, each of which needs to be
investigated in this paper and forthcoming papers (See Fig 1 for the schematic 
representation of the  new scenario).

\subsection{Stage 1: Formation of a massive gaseous clump
in a gas-rich dwarf}
A massive gaseous clump with the initial masses as large as
$10^7-10^8 {\rm M}_{\odot}$ is formed from gravitational instability
of the gaseous disk in a dwarf galaxy. Such a clump corresponds
to a  massive GMC or a GMC complex and has been  demonstrated to be formed
in luminous disk galaxies and dwarfs,
if the gas mass fractions are higher  
(e.g., Shlosman \& Noguchi 1993; Noguchi 1998;  Bekki 2007).
These massive clumps are highly likely to have fractal structures
like the Galactic GMCs, and hierarchical cluster complexes can be formed
from such fractal structures.
These massive clumps can be the progenitor of galactic bulges 
(e.g., Noguchi 1999; Elmegreen et al. 2008) and GCs 
(Shapiro et al. 2010; Adamo et al. 2013; Bekki 2015a, Bekki 2016). 
Star formation (i.e., FG star formation) starts to proceed very efficiently within the clumps,
and consequently hierarchical stellar structures are developed.

\subsection{Stage 2: Formation of a SCC from hierarchical stellar structure within the fractal clump
}

Numerous stellar associations and clusters are  formed from 
dynamical relaxation processes of the hierarchical stellar structure within the clump with initial fractal structures.
The most massive cluster (`the main cluster') 
in these stellar objects is regarded as the progenitor of a
GC  in the new scenario.
The mass function of the SCC  is likely to follow either the mass fraction
of GMCs observed in nearby galaxies
($\beta$ of $1.71-2.49$;  Blitz et al. 2007) 
or that of young star clusters  (e.g., de Grijs et al. 2003).
Some very low-mass unbound clusters and stellar 
associations  could have been already disintegrated
owing to dynamical relaxation process to become field stars in the SCC
at this stage.
Since the mass fraction of such field stars in the SCC can not be 
estimated in the present study without further numerical simulation of 
the SCC formation, we simply assume that such a fraction is zero in the
present simulation.

\subsection{Stage 3: Explosions of multiple SNe and the formation of 
super-giant gaseous hole }

Since the SCC has a large number of massive stars with $m_{\rm s} \ge 8 M_{\odot}$,
energetic feedback effects of massive stars and SNe can ionize cold gas left after star formation
in the clump,
brow out the gas from the host dwarf galaxy, 
and finally develop a giant gas hole with the diameter as large
as 1 kpc (as LMC4 area in the LMC).
The sizes of such giant holes depend on the masses of SCCs  such that
larger gaseous holes can be created in more massive SCCs owing to a larger number of SNe.
Although there can be hot, tenuous gas left in the SCC,  the total mass of such
gas is negligible.

\subsection{Stage 4: Gas ejection from AGB stars}

When massive stars with $m_{\rm s}=8 {\rm M}_{\odot}$ enter into the AGB phases
(roughly 30-40 Myr after their birth; 
$t_{\rm agb}(m_{\rm s}=8 {\rm M}_{\odot}) \approx 30-40$ Myr),
stellar winds from the AGB stars can supply gas for further star formation in the SCC.
There are two
key questions here regarding the validity of the new scenario.
One key question
is whether the SCC is still a collection of clusters  
(i.e., whether the hierarchical structure can survive) when stars with 
$m_{\rm s}=8 {\rm M}_{\odot}$ become AGB stars.
Merging of SCs during SCC formation from fractal gas clumps (GMCs or
GMC associations) has  partially or completely wiped out
the initial hierarchical stellar structures of SCCs
when massive AGB stars start to eject gas.
A physical condition for survival of hierarchical structures
in a SCC is described as follows:
\begin{equation}
t_{\rm merge} >  t_{\rm agb}(m_{\rm s}=8 {\rm M}_{\odot}),
\end{equation}
where $t_{\rm merge}$ is merging timescale of SCs.
This $t_{\rm merge}$ can be shorter
than $\sim 30$ Myr in low-mass SCCs, which means that
their initial hierarchical structures have been at least partially
(or completely)
lost in such low-mass SCCs. Our future simulations need
to investigate to what extent initial hierarchical structures
have been lost at the time of gas ejection from AGB stars
with $m_{\rm s}=8 {\rm M}_{\odot}$ for a given SCC mass.

Bastian et al. (2005) estimated ages of stellar populations in
low-mass SCCs with masses of [3-30]$\times 10^4 {\rm M}_{\odot}$
and found that they are less than $10^7$ yr old (typically a few Myr old).
Their SCCs are quite low-mass ones from  which GCs can not be formed in
the present scenario. 
Murray (2011) showed the lifetimes of the Galactic GMCs with star formation
are only a bit shorter than 3 free-fall times, which means that
the typical lifetime of the GMCs is $\sim 30$ Myr.
If the ages (or age differences of stellar populations)
of SCCs correspond to the lifetimes of their host GMCs,
then it is expected that more massive SCCs (from more massive GMCs)
can have older ages (or larger age spreads).
Accordingly, if the ages of such low-mass SCCs are a few Myr
(or at most 10 Myr), then the ages of high-mass SCCs 
($\sim 10^7 {\rm M}_{\odot}$) should be significantly older than
a few Myr.
It is thus highly likely that hierarchical structures can survive longer
than 30 Myr in  massive SCCs with their masses larger than
$10^7 {\rm M}_{\odot}$. It is our future work to investigate
this issue using hydrodynamical simulations of star-forming GMCs with
fractal structures.

The other related question  is whether the SCC can keep its clustering status
for an enough long time such that gas ejected
from AGB stars evolving from intermediate-mass stars
($3 \le m_{\rm s}/{\rm M}_{\odot} \le 8$) can interact with 
the main cluster and can be subsequently captured by the cluster.
Star clusters in a SCC can merge with one another to
form a massive single cluster 
(Kroupa 1998; Fellhaure \& Kroupa 2002; Bekki et al. 2004b) and the time scale 
for the completion of violent merging is as short as
10 $t_{\rm cr}$, where $t_{\rm cr}$ is the crossing time scale
of the SCC  (Fellhaure \& Kroupa 2002; Bekki et al. 2004b).
This merging timescale $t_{\rm merge}$
corresponds to  $70-380$ Myr for SCCs with a mass
of $2 \times 10^7 {\rm M}_{\odot}$ and  sizes of $50-150$ pc
(Fellhaure \& Kroupa 2002).
In the present scenario, $t_{\rm merge}$ is as follows:
\begin{equation}
t_{\rm merge}= 3.9 \times 10^8 { ( \frac{\sigma  }{ 10 {\rm km s^{-1} } } ) }^{-1}
 ( \frac{ r_{\rm scc} }{ 200 {\rm pc} } ) ,
\end{equation}
where $\sigma$ is the stellar velocity dispersion of the SCC-host dwarf at
the location of the SCC and $r_{\rm scc}$ is the radius of the SCC with
a spherical cluster distribution. Here the relative velocity between the SCC
and a member SC is assumed to be the same as $\sigma$. 
The following relation
is therefore ensured for very massive SCCs:
\begin{equation}
t_{\rm merge} >  t_{\rm agb}(m_{\rm s}=[3-8] {\rm M}_{\odot}).
\end{equation}
This means that gas from AGB stars that evolved from
stars with $m_{\rm s}=[3-8] {\rm M}_{\odot}$ can interact with the main cluster
well before the SCC becomes dynamically relaxed through violent merging.
It is likely that low-mass clusters are stripped before they merge with the main cluster
owing to the tidal filed of the host dwarf. 
Also, low-mass SCCs might have been well relaxed when gas from AGB stars is being
accreted onto the central regions of SCCs.
These points can be investigated
in the present numerical simulations.
Just for convenience,  time $T$ in a simulation  is set to be 0 at  this stage 4.

\subsection{Stage 5:  Accretion of AGB ejecta and cold gas  onto the main cluster
within the SCC
}

Intermediate-mass stars  in stellar associations and low-mass clusters
start to eject gas into ISM of the dwarf   through stellar winds.  A significant
fraction of the AGB ejecta
can be smoothly accreted onto the main cluster owing to the lack of hot gas in the hole.
However, some fraction of low-mass clusters can be quickly stripped from the main cluster
by the tidal field of the host dwarf so that most of
their AGB ejecta can not be accreted onto the main cluster.
Low-mass clusters that merge with the main cluster can provide more AGB ejecta
for the main cluster. 
Cold gas initially outside the gas hole 
 is not chemically polluted by the ejecta from SNe of the SCC
and thus has chemical abundances similar to those of the main cluster.
Accretion of the cold gas onto the main cluster
comes later than that of the AGB ejecta in this scenario. 
Such cold gas accretion can dilute the AGB ejecta, and consequently, the chemical abundances
of the mixed gas can be quite different from those of the AGB ejecta.

\subsection{Stage 6:  Secondary star formation from the accreted gas
}

Secondary star formation from the accreted gas starts whenever the physical properties of the gas
satisfy the physical conditions required for star formation in
dense stellar systems. Therefore, it is possible that AGB ejecta is converted into
new stars without mixing with cold gas (i.e., well before the cold gas accretion onto the
main cluster) in some cases. If this star formation lacks massive 
with $m_{\rm s} \ge  8 M_{\odot}$,
then it can continue more than a few Myr without being influenced by energetic SN feedback
effects. The IMF of the formation of SG stars  is a key factor that determines
the duration of secondary star  formation (and gas accretion)
thus the total mass of SG stars.

\subsection{Stage 7:  Complete disintegration of the SCC 
}

Star formation can continue  as long as  accretion of AGB ejecta and
cold gas  onto the main cluster continues.
Accretion of gas ejected from intermediate-mass AGB stars
almost completely stops when the original SCC is  disintegrated
 by the tidal field
of the host dwarf.  The tidal radius ($r_{\rm t}$) of a SCC in
a dwarf galaxy is estimated as follows:
\begin{equation}
r_{\rm t}= 180 ( \frac{ M_{\rm scc} }{ 10^7 {\rm M}_{\odot} } )^{1/3}
( \frac{ v_{\rm c} }{ 60 {\rm km s^{-1} } } )^{-2/3}
( \frac{ R_{\rm scc} }{ 1 {\rm kpc} }  )^{2/3} pc,
\end{equation}
where $v_{\rm c}$ is the circular velocity of the host dwarf
at the position of the SCC. Accordingly,
low-mass SCs initially 
outside $r_{\rm t}$ are quickly  stripped by the host dwarf
so that their AGB stars
can not contribute to the gas accretion onto the main clusters.
Also, if the host dwarf galaxy is destroyed by a large
galaxy, then cold gas accretion on the main cluster can be shut down too.
Some low-mass SCs can merge with the main cluster before the disintegration of the SCC.
They are destroyed during merging to from a stellar halo around the main cluster,
and only a small fraction of their stars can be within the effective radius of the main
cluster. 

\begin{figure}
\psfig{file=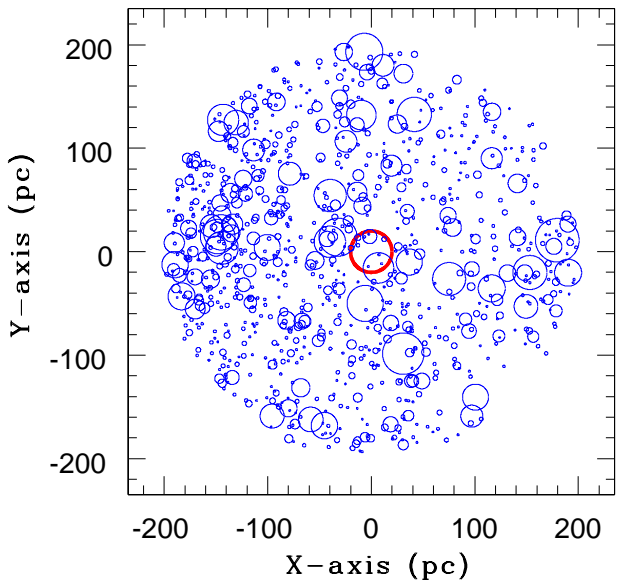,width=8.5cm}
\caption{
Spatial distribution of low-mass star clusters (SCs)  around the main cluster
in the fiducial model. Each blue circle represents the location of a SC
with the size indicating  the  mass of the SC. The red circle represents
the size of the main cluster. Since there are too many very low-mass
SCs in the SCC,  only 10\% of the SCs are shown for clarity.
In the present study, it is assumed that most SCs still exist 
(i.e., without being disintegrated) in a SCC 
when AGB stars start to eject gas (i.e., 30-40 Myr after SC formation).
However, this assumption can be less realistic, because merging of SCs 
and destruction of small SCs have been ongoing since the formation
of SCs from a fractal  SCC-hosing GMC. This assumption is adopted
just for the purpose of estimating gas accretion rates onto forming GCs
in the present GC formation scenario.
Real GC formation processes can be more complicated than the present
model.
}
\label{Figure. 2}
\end{figure}

\begin{table*}
\centering
\begin{minipage}{160mm}
\caption{The basic model parameters for
SCCs.}
\begin{tabular}{llllllll}
{ Model ID } &
{ Dwarf type \footnote{ The parameter values for each dwarf model is given
in Table 1.
 }} & 
{ $m_{\rm mc}$ \footnote{ The initial total mass for the main cluster
in units of $10^{6} M_{\odot}$.
 }} & 
{ $r_{\rm mc}$  \footnote{  The initial size of the main cluster
in units of pc. The scale length ($a_{\rm mc}$)  is roughly $0.2r_{\rm mc}$
}} &
{ $R_{\rm scc}$  \footnote{ The initial distance of the SCC
from the dwarf galaxy's center in units of pc. For high-$z$ dwarf models,
$R_{\rm scc}$ is chosen (=0.57) such that $R_{\rm scc}/R_{\rm s}$ is the  same
between high-$z$ and other models with $R_{\rm scc}=1$ kpc.
}} &
{ $f_{\rm agb}$  \footnote{ The mass fraction of AGB ejecta to the initial
mass of a SC ($m_{\rm sc}$).
}} &
{ $r_{\rm scc}$  \footnote{ The initial size of a SCC
in units of pc.
}} & 
{ comments } \\
M1 & DW1 & 0.5 & 20 & 1.0 & 0.05 & 200 & fiducial \\
M2 & DW1 & 0.2 & 20 & 1.0 & 0.05 & 200 &  \\
M3 & DW1 & 1.0 & 20 & 1.0 & 0.05 & 200 &  \\
M4 & DW1 & 0.5 & 20 & 0.01 & 0.05 & 200 & nuclear SC model \\
M5 & DW1 & 0.5 & 20 & 0.3 & 0.05 & 200 &  \\
M6 & DW1 & 0.5 & 10 & 1.0 & 0.05 & 200 &  \\
M7 & DW1 & 0.5 & 50 & 1.0 & 0.05 & 200 &  \\
M8 & DW1 & 0.5 & 20 & 1.0 & 0.02 & 200 &  \\
M9 & DW1 & 0.5 & 20 & 1.0 & 0.1 & 200 &  a more top-heavy IMF\\
M10 & DW2 & 0.5 & 20 & 1.0 & 0.05 & 200 &  \\
M11 & DW6 & 0.5 & 20 & 0.57 & 0.05 & 200 &  \\
M12 & DW1 & 0.5 & 20 & 1.0 & 0.05 & 200 & 
$1.2 \times 10^2 \le m_{\rm sc}/{\rm M}_{\odot} \le 3.6 \times 10^3$  \\
M13 & DW1 & 0.5 & 20 & 1.0 & 0.05 & 200 & 
$1.2 \times 10^2 \le m_{\rm sc}/{\rm M}_{\odot} \le 1.2 \times 10^5$  \\
M14 & DW1 & 0.5 & 20 & 1.0 & 0.05 & 200 & 
$\beta=0$ ($m_{\rm sc}=1.2 \times 10^3 {\rm M}_{\odot}$  for all SCs)\\
M15 & DW1 & 0.5 & 20 & 1.0 & 0.05 & 100 &  \\
M16 & DW1 & 0.1 & 10 & 1.0 & 0.05 & 200 &  \\
M17 & DW1 & 0.03 & 20 & 1.0 & 0.05 & 200 &  \\
M18 & DW1 & 0.03 & 20 & 1.0 & 0.05 & 200 &  
$3.6 \times 10 \le m_{\rm sc}/{\rm M}_{\odot} \le 1.2 \times 10^3$  \\
M19 & DW4 & 0.5 & 20 & 1.0 & 0.05 & 200 & Fiducial with cold ISM \\
M20 & DW4 & 0.2 & 20 & 1.0 & 0.05 & 200 &  \\
M21 & DW4 & 1.0 & 20 & 1.0 & 0.05 & 200 &  \\
M22 & DW4 & 0.5 & 20 & 1.0 & 0.05 & 200 & a smaller gas hole with $r_{\rm h}=200$ pc \\
M23 & DW5 & 0.5 & 20 & 1.0 & 0.05 & 200 & gas-poor host dwarf \\
M24 & DW7 & 0.5 & 20 & 0.57 & 0.05 & 200 & high-z more compact model  \\
M25 & DW3 & 0.5 & 20 & 1.0 & 0.05 & 200 & low stellar density of the dwarf  \\
M26 & DW4 & 0.5 & 20 & 1.0 & 0.05 & 200 &  \\
M27 & DW8 & 0.5 & 20 & 0.57 & 0.05 & 200 & high-Z, lower stellar density of the dwarf   \\
\end{tabular}
\end{minipage}
\end{table*}

\section{The model}

We focus exclusively on the Stage 4 and 5 in this paper and thereby
investigate the total mass of gas accreted onto main clusters orbiting
around their host dwarf galaxies.
We will investigate
physical processes in  other stages of the new GC formation scenario in our future papers. 
In order to investigate the accretion processes of AGB ejecta and cold ISM,
we use our original simulation code (Bekki 2013; Bekki 2015b, c) that can be
run on GPU clusters. The code enables us to investigate chemical evolution,
dust formation and evolution (Bekki 2013),  formation of molecular hydrogen on
dust grains (Bekki 2015b),  photo-electric heating of gas by
dust and star formation in galaxies (Bekki 2015c). Since the details of the code
are already given in our previous papers, we just briefly explain the code
in the present study.  The time $T=0$ in each simulation
corresponds to when massive AGB stars ($m_{\rm s}=8 {\rm M}_{\odot}$)
start to eject gas into ISM of dwarf galaxies through stellar winds.

\subsection{Dwarf disk galaxy}

We consider a gas-rich dwarf disk galaxy as a host of a SCC in the present study.
The gas-rich dwarf 
consists of a dark matter halo,
a stellar disk,  a gas disk, and a SCC.
The dark matter halo, the stellar disk,
are all represented by collisionless N-body particles.
Hydrodynamics of the gas disk,  conversion from gas into
new stars (`star formation'),  chemical evolution,
and dust formation and evolution are all included in the present study.
However, we do not include some dust-related physical processes,
such as photo-electric heating and gas-dust drag in this study,
because the main purpose of this paper is not to discuss such dust effects on
galaxy evolution. The total masses of these components
are denoted as $M_{\rm h}$, $M_{\rm s}$, $M_{\rm g}$, and $m_{\rm scc}$,
respectively.
The dark matter halo
has the `NFW' one (Navarro et al. 1996) density profile
with a central cusp predicted by the Cold Dark Matter (CDM)  model:
\begin{equation}
{\rho}(r)=\frac{\rho_{0}}{(r/r_{\rm s})(1+r/r_{\rm s})^2},
\end{equation}
where $r$,  $\rho_{0}$,  and $r_{\rm s}$ are the distance from the center
of the cluster, the central density, and the scale-length of the dark halo,
respectively.
We investigate only the models with $M_{\rm h}=10^{10} {\rm M}_{\odot}$
in the present study.
The virial radius ($r_{\rm vir}$),  the scale radius ($r_{\rm s}$),
and the `$c$' parameter (=$r_{\rm vir}/r_{\rm s}$)
are chosen such that the values
are consistent with recent cosmological simulations
for the adopted $M_{\rm h}$
(Neto et al. 2007).

The radial ($R$) and vertical ($Z$) density profiles of the stellar 
and gaseous  disks of  a dwarf are
assumed to be proportional to $\exp (-R/R_{0}) $ with scale
length $R_{0} = 0.2R_{\rm s}$ and to ${\rm sech}^2 (Z/Z_{0})$ with scale
length $Z_{0} = 0.04R_{\rm s}$ , respectively.
Although the gas mass fraction ($f_{\rm g}=M_{\rm g}/M_{\rm s}$) 
is assumed to be a free parameter
($0 \le f_{\rm g} \le 1$),
we mainly investigate the models with $f_{\rm g}=0$ (i.e., `without cold ISM'),
because we intend to understand the accretion of AGB ejecta onto
main clusters in SCCs  more clearly.
In addition to the
rotational velocity caused by the gravitational field of disk
and dark halo components, the initial radial and azimuthal
velocity dispersions are assigned to the disc component according to
the epicyclic theory with Toomre's parameter $Q$ = 1.5.  The
vertical velocity dispersion at a given radius is set to be 0.5
times as large as the radial velocity dispersion at that point.

Star formation from gas in a dwarf galaxy is included as follows.
We assume that the following three physical conditions need to be met
for each gas particle to be converted into a new star.
First  is that the local density ($\rho_{\rm g}$) exceeds a threshold density
(${\rho}_{\rm th}$) for
star formation:
\begin{equation}
{\rho}_{\rm g} > {\rho}_{\rm th}, 
\end{equation}
where $\rho_{\rm th}$ is set to be 100  H atoms cm$^{-3}$.
Second is that the local dynamical time scale is shorter
than the sound crossing time scale, which mimics
the Jeans instability in the gas disk.
Third  is that  the local velocity
field is consistent with that for gravitationally collapsing
(i.e., div {\bf v}$<0$).
Gas mass is assumed to be consumed by star formation according
to the Kennicutt-Schmidt law (Kennicutt 1998).
The power-law slope ($\alpha_{\rm sf}$) of the  Kennicutt-Schmidt law
(SFR$\propto \rho_{\rm g}^{\alpha_{\rm sf}}$)
is set to be 1.5  in the present
study.
It should be noted here that although this 
star-formation model  does a good job 
in predicting
galaxy-wide star-formation, this might not be appropriate for star formation
within  star clusters. Thus we do not discuss much about secondary
star formation within GCs in the present paper.

Initial gaseous metallicities of gas particles in a dwarf
are set to be $-1.6$ for
all models, and there is no   radial metallicity gradient in the dwarf.
Metallicity-dependent radiative cooling is self-consistently modeled
according  to the metallicities of gas particles.
Chemical enrichment processes, dust formation and evolution in
ISM,  and SN feedback effects
are included in the same way as done in our previous
simulations (Bekki 2013; Bekki 15b,c).
However, the details of these modeling are not so important in the present
simulations, because we mainly investigate gas accretion onto massive
SCs only for $\sim 300$ Myr.

We investigate dwarf galaxy models with
a fixed $M_{\rm h}=10^{10} {\rm M}_{\odot}$ yet different
$M_{\rm s}$ and $f_{\rm g}$.  
The eight dwarf models are investigated, and the parameter values
are given in Table 1.
The total number of particles for dark matter, stars,
gas (inclusive of AGB ejecta), and a main cluster
in a model  are 500000, 500000, 150000,
and 10000, respectively. These four components
have different initial  gravitational softening lengths ($\epsilon$)
according
to their initial half-number radius for
each component. For example,  $\epsilon$ is set to be
194pc for dark matter,  14.8 pc for stars and gas,
and 0.47 pc for the main cluster in the dwarf  model
DW1. The high-z dwarf models (DW6, 7, and 8) have more compact
distributions of baryon and dark matter, and the mean density of
dark matter halo is consistent with that 
expected for  dwarf galaxies with $M_{\rm h}=10^{10} {\rm M}_{\odot}$
at $z=2$.

\subsection{SCC}

A SCC is assumed to consist of one main cluster represented by
the Plummer model with a size of $r_{\rm mc}$
and a scale length of $0.2 r_{\rm mc}$  and numerous low-mass clusters
by point-mass particles. 
It would be  possible that one SCC has a few  massive
star clusters ($m_{\rm mc}>10^5 {\rm M}_{\odot}$) ,
we here investigate a case where one SCC has only one massive main
cluster, because we can more clearly understand the roles of other clusters 
in the gas accretion on main clusters in SCCs  without introducing other model parameters.
The SCC has a power-law cluster mass function
as follows,
\begin{equation}
N(m_{\rm sc})= N_0 m_{\rm sc}^{-\beta}
\end{equation}
where $\beta=2$, which is expected from hierarhcial star formation
(e.g., Elmegreen 2008 for a review) and $N_0$ is a constant. 
The mass of the main cluster 
is a free parameter denoted by $m_{\rm mc}$,
and the lower and upper  mass cut-offs are denoted by
$m_{\rm l}$ and $m_{\rm u}$, respectively.
The values of 
$m_{\rm l}$ and $m_{\rm u}$
are set to be 
$1.2 \times 10^3 {\rm M}_{\odot}$ and 
$3.6 \times 10^4 {\rm M}_{\odot}$,
respectively,  for most models in the present study.
Low-mass clusters in the SCC are distributed within a sphere
with a radius $r_{\rm scc}$ (= SCC size). 
The initial 3D velocities of each cluster is chosen such that they
can be the same as those of a field star (of the host
dwarf galaxy) that is closest to the cluster.
The initial distribution of SCs within $r_{\rm scc}=200$ pc is shown
in Fig. 2. 

The SCC initially embedded in a giant gas hole is assumed
to  orbit  around the center of 
a  dwarf galaxy. The SCC is initially located at ($x$, $y$, $z$) = ($R_{\rm scc}$, 0, 0),
where    
the 3D coordinate of the host dwarf's center  is set to be (0, 0, 0)
and $R_{\rm scc}$ is a parameter that controls the initial distance of the SCC from
the dwarf's center.
The SCC is assumed to have a circular orbit within the dwarf disk
and the circular velocity is determined by the mass distribution of the dwarf.
The gas hole is assumed to be created by energetic feedback effects of SNe of the SCC itself.
Such a giant gas hole is observed in the LMC (e.g., LMC 4) and 
could have been formed as a results of energetic feedback effects
such as SNe explosion. 
The gas hole is assumed to have a circular shape and its radius is a free parameter
defined by $r_{\rm h}$. No cold gas of the host dwarf galaxy
is assumed to exist initially within $r_{\rm h}$.

We mainly investigate the evolution of SCCs with the initial stellar masses
($M_{\rm scc}$) of $\sim 10^7 {\rm M}_{\odot}$, 
because more than $10^5 {\rm M}_{\odot}$ gas can be 
accumulated in such massive SCCs.
The initial masses of gas clumps
($M_{\rm clump}$)  hosting such SCCs
are described as follows:
\begin{equation}
M_{\rm clump}=10^8
{ ( \frac{ \epsilon_{\rm scc} }{0.1} ) }^{-1}
( \frac{ M_{\rm scc} }{ 10^7 {\rm M}_{\odot} } )
{\rm M}_{\odot},
\end{equation}
where $\epsilon_{\rm scc}$ is the formation efficiency of a SCC within
a gas clump.
If $\epsilon_{\rm scc}$ corresponds to 
cluster formation efficiency (CFE), then
the original gas  mass of
a GC with multiple stellar populations
can be  $\sim 10^8 {\rm M}_{\odot}$ for a reasonable
CFE of 0.1 
(e.g., Adamo et al. 2015; Johnson et al. 2016).
Although
this means that the original gas clumps should be rather massive,
the adopted low CFE of 0.1 does not meet the physical conditions
(i.e., CFE significantly higher than 0.1)
required for the formation
of bound SCs (e.g., Hills 1980).
Accordingly,
the original clump masses of GCs can be significantly lower
than the above  $\sim 10^8 {\rm M}_{\odot}$.

The present SCC model is more realistic than previous GC formation models
with a single massive cluster, and it allows us to investigate
the roles of hierarchical star cluster complexes in GC formation.
However, the model 
is still less realistic 
at some points (i.e., central massive clusters within SCCs)
and needs to be improved in our future works. 
It should be stressed that the present model is chosen such that
gas accretion processes onto proto-GCs can be investigated in
a quantitative manner. Real cluster formation processes from SCCs are more complicated
than the present model describes. For example, the observed
large fraction of binary clusters in the LMC (e.g., Bhatia \& Hatzidimitriou 1988)
can not be simply explained by the present model, and it can be better explained
by other cluster formation models based on GMC collisions (e.g., Bekki et al. 2004a).
Thus we need to discuss these other issues related to cluster formation in our
future papers.

\begin{figure}
\psfig{file=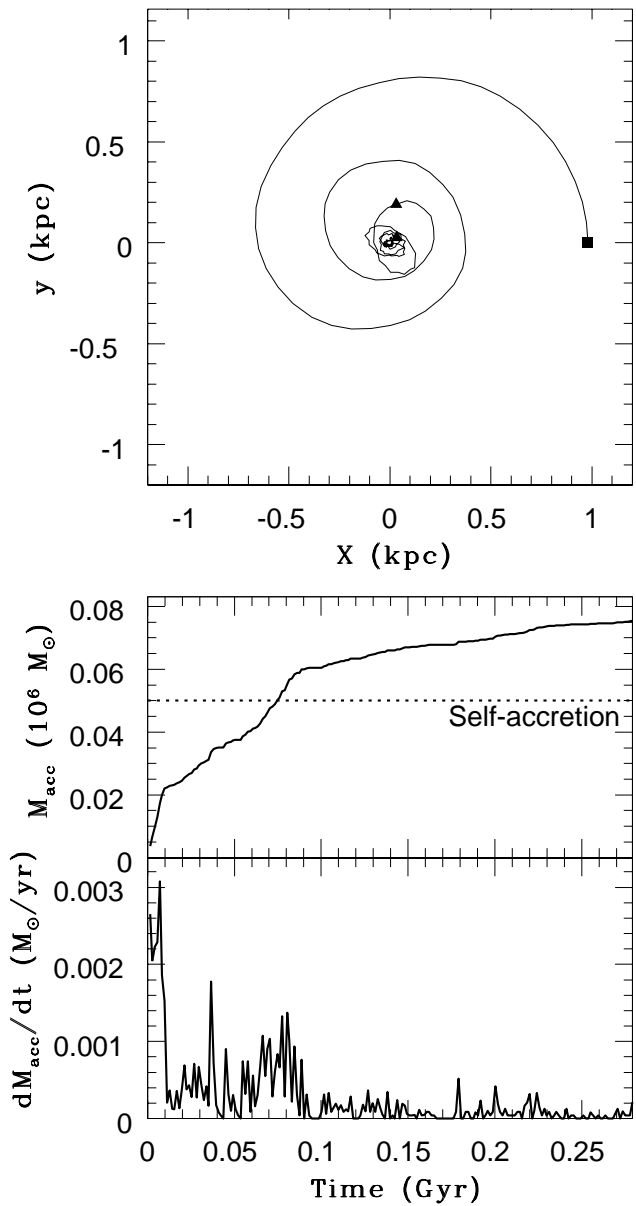,width=8.5cm}
\caption{
Time evolution of the orbit of the 
main cluster projected onto the $x$-$y$ plane (top),
the total gas mass accreted onto the main cluster
($M_{\rm acc}$; middle), and the accretion rate ($dM_{\rm acc}/dt$; bottom) 
for the fiducial model M1. The filled square indicates  the initial
position of the main cluster and the two filled triangles 
indicate the location of the cluster at $T=0.1$ and 0.2 Gyr in
the top panel. The main cluster sinks into the inner region owing
to dynamical friction of the cluster against the disk field stars.
 The dotted line in the middle panel
indicates the maximum possible gas mass  that can be accreted
onto the main cluster from AGB stars of the cluster itself
(`self-accretion').
}
\label{Figure. 3}
\end{figure}

\subsection{Ejection and accretion of gas from AGB stars}

The wind velocity ($v_{\rm w}$)  of gas ejected from AGB stars is assumed to be 10 km s$^{-1}$,
which is consistent with the observed value for low-metallicity
 AGB stars in the LMC 
(e.g., Marshall et al. 2004).
A gas particle (`AGB particle') is ejected with $v_{\rm w}$ from each  SC particle only once 
and the initial total mass of the gas particle is $f_{\rm agb} m_{\rm sc}$,
where $f_{\rm agb}$ is the mass fraction of AGB ejecta and $m_{\rm sc}$ is the initial
total mass of the SC. 
This $f_{\rm agb}$ depends on the initial mass function (IMF) of stars
(B11). In the present study, the IMF for a stellar system  is assumed to have  a power-law
function with a slope of $\alpha$  
as follows:
\begin{equation}
\psi (m_{\rm s}) = M_{\rm 0} m_{\rm s}^{-\alpha},
\end{equation}
where $M_{\rm 0}$ is a constant derived from the total
mass of the stellar system  and $m_{\rm s}$ is a stellar mass 
($0.1 \le \frac{ m_{\rm s} }{ {\rm M}_{\odot} } \le 100$)  and
$\alpha=2.35$ corresponds to the canonical (Salpeter) IMF.
We adopt the following relation between the initial
stellar  mass ($m_{\rm s, i}$)
and the total mass of gas ejected
from the star ($m_{\rm s, ej}$) (Weidemann 2000; B11):
\begin{equation}
m_{\rm s, ej} =  0.916m_{\rm s, i}-0.444.
\end{equation}
For the canonical IMF with $\alpha=2.35$,
$f_{\rm agb}$ can be as large as 0.1 about 300  Myr after the initial burst
of star formation in GC formation.  
Furthermore, $f_{\rm agb}$ can be larger for more top-heavy IMF with
$\alpha$ smaller than 2.35 (See Fig 1 in B11).
Since we investigate the evolution of SCCs only for $\sim 300$ Myr
in the present study,  it is appropriate to choose $f_{\rm agb}$ for stars that
can commence their AGB phases within 300 Myr. Guided by the results of $f_{\rm agb}$ evolution
shown in B11,  we mainly investigate the models with $f_{\rm agb}=0.05$.
The results of other models with $f_{\rm agb}$=0.1 and 0.03 are briefly discussed.

It is highly unrealistic to assume that only one SCC is formed in a dwarf galaxy
at a given time. Accordingly, there should be many  AGB stars evolving
from field stars that form almost simultaneously with the SCC. We therefore
assume that 10\% of field stars are also ejecting gas through AGB winds.
These AGB ejecta from field stellar populations can not contribute
significantly to the gas accretion onto the main cluster of the SCC: we have
confirmed this through a comparative experiment in which no field AGB stars
are included. However, it should be stressed that AGB ejecta from field stellar
populations can interact with AGB wind of the SCC. Therefore, the dynamical
evolution of AGB ejecta from the SCC can be slightly influenced by
the AGB ejecta from the field stellar populations of the SCC-hosting dwarf.

In order to estimate the time evolution of the accretion rate
($ {\dot{M}}_{\rm acc}$)
of AGB ejecta for the main
cluster of a SCC in a simulation, 
we count the number of AGB particles that are within the cluster radius
($r_{\rm mc}$) and have velocities less than 
the escape velocity of the cluster.
Accordingly, we estimate the relative velocity of
a AGB particle with respect to the main cluster ($v_{\rm rel}$)
at each time step.
The AGB ejecta is regarded as being accreted onto the main cluster,
if it meets the following condition:
\begin{equation}
v_{\rm rel} < v_{\rm esc}(\phi (r, t)),
\end{equation}
where $v_{\rm esc}$ is the escape velocity of the main cluster,
$\phi (r,t)$ is the gravitational potential of the main cluster
at the distance $r$ from the cluster's center at time $t$,
$\phi$ is dependent on time and place 
 owing to the mass loss by the tidal field of
the host dwarf,
and $r$ is the distance of the accreted AGB ejecta (i.e., $r<r_{\rm mc}$)
from the center of the main cluster. By estimating
both $v_{\rm rel}$ and $\phi$ at each time step, we avoid counting
AGB particles that happen to be within $r_{\rm mc}$ yet are not
gravitationally trapped by the cluster.

We assume that these AGB particles gravitationally trapped by the main cluster can be
finally used for secondary star formation within the cluster.
The total mass of gas accreted on the main cluster ($M_{\rm acc}$)
is estimated as follows:
\begin{equation}
M_{\rm acc}(t)=  \sum_{i=1}^{nstep} {\dot{M}}_{\rm acc, \it i} dt_i
\end{equation}
where $M_{\rm acc, \it i}$ ($dM_{\rm acc}/dt$) is the 
gas accretion rate, $dt_i$ is the time step width,
and nstep is the number of time steps 
for which $M_{\rm acc}$ is estimated in each simulation. 
Once AGB particles are gravitationally trapped by the main cluster
and its accretion rate is estimated at $T=t_i$,
then these particles are not used for the estimation of gas accretion
rates in the following time steps ($T> t_i$).
The total mass of the accreted gas
is either from AGB ($M_{\rm acc, agb}$)
or  from ISM ($M_{\rm acc, ism}$) in the present study,  and they
can be separately estimated:
\begin{equation}
M_{\rm acc}=M_{\rm acc, agb}+M_{\rm acc, ism} .
\end{equation}
It is possible that not only AGB stars initially in the main cluster
and other low-mass clusters of a SCC but also  field AGB stars 
that form simultaneously with the SCC can contribute to $M_{\rm acc, agb}$.
Therefore, $M_{\rm acc, agb}$ is further described by three terms as follows:
\begin{equation}
M_{\rm acc, agb}=M_{\rm acc, agb, mc}+M_{\rm acc, agb, sc}+M_{\rm acc, agb, f},
\end{equation}
where  $M_{\rm acc, agb, mc}$, $M_{\rm acc, agb, sc}$, and  $M_{\rm acc, agb, f}$
are the accreted AGB ejecta from the main cluster of a SCC,
low-mass clusters of the SCC, and field AGB stars, respectively.
The contribution of field AGB stars is very minor in the present study.
Bekki \& Mackey (2009) and 
Pflamm-Altenburg \&  Kroupa (2009)
investigated how star clusters can capture cold molecular gas
or accrete ISM in galaxies using idealized modeling of the accretion process.
The present work and theirs are therefore  complementary to each other.

\subsection{A parameter study}

We first investigate the models without cold ISM of  gas-rich dwarf galaxies
in order to show more clearly the physical process of the accretion of AGB ejecta onto
the main clusters. We then investigate the models with cold gas of dwarf galaxies
in order to estimate the total gas accretion rates. Since the key parameters in the present study
are $m_{\rm mc}$, $R_{\rm scc}$, and $f_{\rm agb}$, we mainly discuss how
the present results depend on these parameters. We only briefly discuss the importance
of other parameters (e.g., $r_{\rm mc}$ and dwarf structures) in the gas accretion
processes in the present study. The values of the model parameters are summarized
for each model in Table 2.

\begin{figure}
\psfig{file=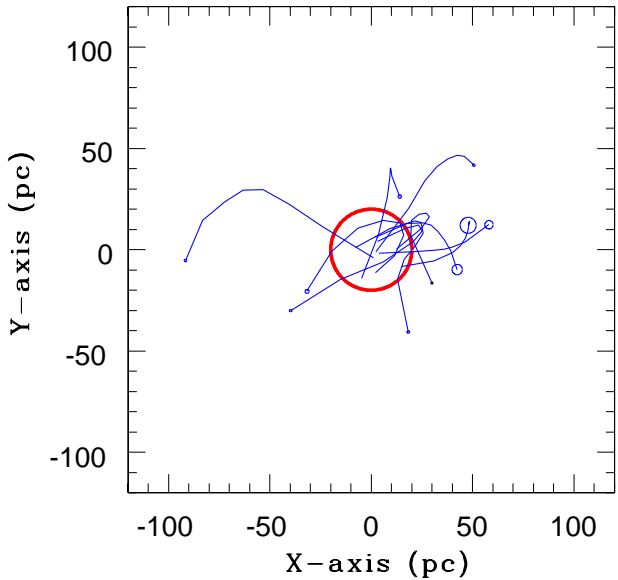,width=8.5cm}
\caption{
Orbital evolution of selected ten AGB particles that are accreted onto the main cluster
at $T=0.014$ Gyr (blue solid lines) in the fiducial model M1. These orbits are 
projected onto the $x$-$y$ plane (i.e., the disk plane of the host dwarf).
The thick red line indicates the accretion
radius ($=r_{\rm mc}$) of the main cluster. The initial masses of SCs
from which AGB particles are ejected  are indicated
by the sizes of blue circles. Clearly, the AGB particles come from different areas of
the SCC and have different orbital angular momentum with respect to the main cluster.
}
\label{Figure. 4}
\end{figure}

\begin{figure}
\psfig{file=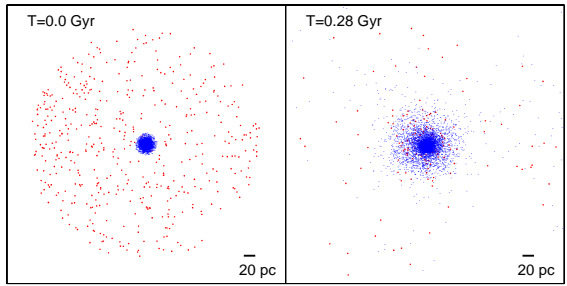,width=8.5cm}
\caption{
Initial distributions of SCs (red) and stars of the main cluster (blue)
within the SCC projected onto
the $x$-$y$ plane at $T=0$ (left) and $T=0.28$ Gyr (right) in the fiducial model.
For consistency with Fig. 2,
only 10\% of the low-mass SCs  are shown.
}
\label{Figure. 5}
\end{figure}

\section{Results}

\subsection{Without cold ISM}

\subsubsection{Gas accretion from other clusters}

Fig. 3 shows that while the main cluster spirals into the central
region of the host dwarf owing to dynamical friction,
the main cluster can accrete AGB ejecta steadily from
other low-mass SCs initially located in its surrounding
in the fiducial model M1. 
The derived short timescale of dynamical friction 
of the main cluster appears to be at odds with
the mass of just $m_{\rm mc}=5 \times 10^5 {\rm M}_{\odot}$.
However, this rapid spiraling-in can be understood in terms of
much more efficient dynamical friction of the SCC itself  against disk field stars
of the dwarf. The SCC with a mass of $\sim 10^7 {\rm M}_{\odot}$ can rapidly spiral
into the nuclear region of the dwarf as long as it is not disintegrated by
the dwarf's tidal field. The main cluster can follow the orbit of the SCC so that
the timescale of the main cluster's spiraling-in can be quite short.
Although AGB ejecta from low-mass clusters
can not be gravitationally trapped  by themselves  
owing to their low escape velocities ($<10$ km s$^{-1}$), 
it can be trapped within  the SCC with a much deeper gravitational potential.
Therefore, the following condition is met in the SCC:
\begin{equation}
v_{\rm esc, scc} > v_{\rm w} > v_{\rm esc, sc},
\end{equation}
where $v_{\rm w}$ is the wind velocity of AGB stars,
 $v_{\rm esc, scc}$ and $v_{\rm esc, sc}$ are the escape velocities
of the SCC and its member SCs, respectively.
A significant fraction of the AGB ejecta could be  be finally accreted on
the main cluster in any SCC that meets the above condition in the new scenario.

The accretion rate is higher in  the early evolution
of the SCC with the maximum rate of $\sim 0.003 {\rm M}_{\odot}$ yr$^{-1}$,
mainly because AGB ejecta from 
numerous low-mass SCs that were born within 50pc from the main cluster
can be efficiently and rapidly accreted onto the cluster. 
The total amount of AGB ejecta accreted onto the main cluster from other SCs
can be as large as $8.0 \times 10^4 {\rm M}_{\odot}$, which is significantly
larger than the maximum possible mass of gas ($0.05 m_{\rm mc}$) 
that can be accreted onto the cluster
from the cluster's own AGB stars.
If all AGB ejecta from this massive main cluster 
with $m_{\rm mc} \ge 5 \times 10^5 {\rm M}_{\odot}$ can be accreted onto
the central region of the cluster (which is highly likely),
the total mass of AGB ejecta ($M_{\rm acc}$) 
accreted onto the main cluster within 0.28  Gyr can be
as large as  $1.3 \times 10^5 {\rm M}_{\odot}$ in this model.
This $M_{\rm acc}$ is equivalent to the observed 
total  mass of SG stars  in typical GCs (C09).

Fig. 4 shows the orbits of ten selected AGB particles that are accreted onto the main cluster
by $T=0.014$ Gyr in the fiducial model. 
Clearly, the accreted AGB stars originate from different directions with different
velocities and angular momentum with respect to the main cluster.
Accordingly, if these gas components collide with one another within the main cluster,
then they can lose a large amount of their kinetic energy through
gaseous dissipation owing to their large velocity differences.
These gas accretion processes in the new scenario appear to be quite different
from those described in previous simulations (Bekki 2010 and B11)
in which AGB ejecta can rapidly form a rotating  gas disk
within existing massive SCs
without much energy dissipation.
Such a difference would end up with different histories of star formation (i.e.,
SG star formation) within SCs, which will need to 
be investigated in our forthcoming papers.

Fig. 5 demonstrates that most of the initial SCs 
(71\%) can be stripped from the surrounding of the main
cluster (i.e., outside 200 pc from the cluster)
to become  isolated SCs in the disk of the host dwarf within 0.28 Gyr.
This means that
although AGB ejecta from some SCs can be accreted onto the main cluster
before the SCs are stripped,
other SCs simply lose their AGB ejecta to the field of the host galaxy.
Interestingly, SCs with the total mass of $3.3 \times 10^4 {\rm M}_{\odot}$ can
remain within 5 pc from the cluster at $T=0.28$ Gyr. These SCs 
merged with the cluster during the dynamical
evolution of the SCC so that they can finally  become a part of the cluster.
The mass increase of the main cluster
due to this merging is not so significant in this fiducial model.

\begin{figure*}
\psfig{file=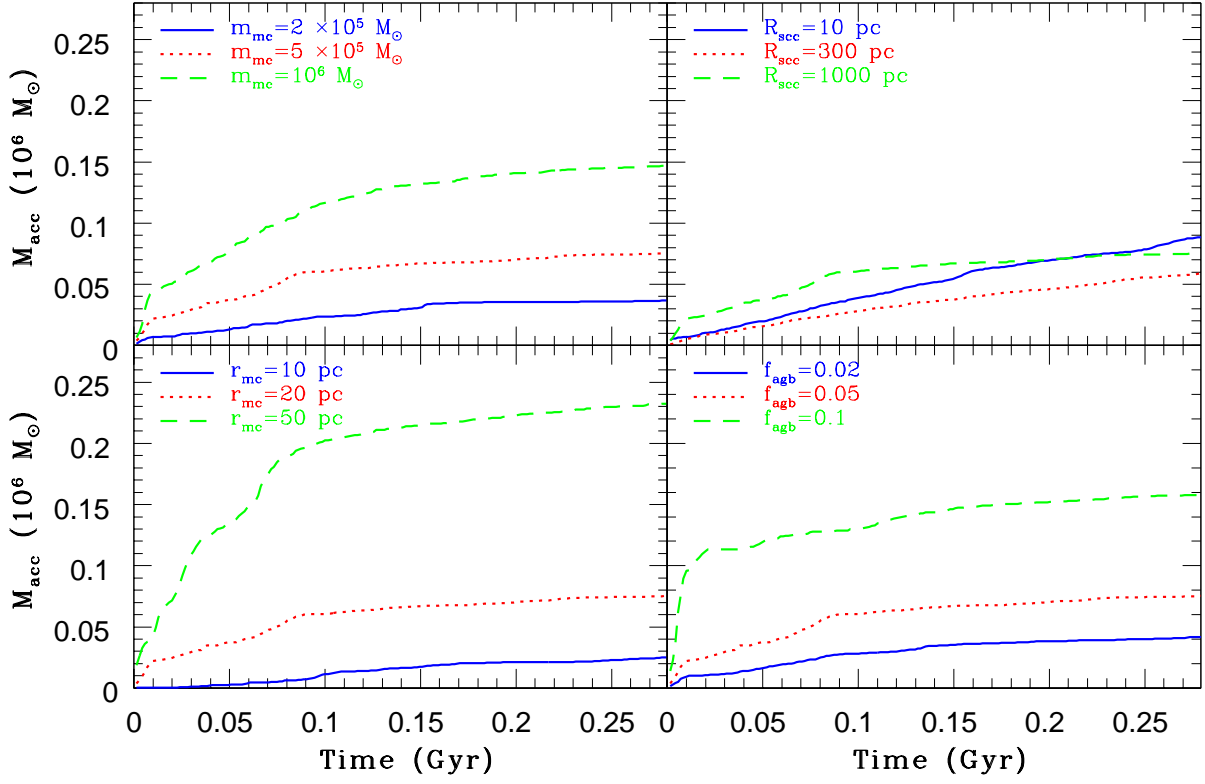,width=16.0cm}
\caption{
Total mass of AGB ejected accreted onto the main cluster ($M_{\rm acc}$) for
different models without cold ISM: different $m_{\rm mc}$ (M1, M2, M3; upper left),
$R_{\rm scc}$ (M1, M4, M5; upper right),
$r_{\rm mc}$ (M1, M6, M7; lower left),
and $f_{\rm agb}$ (M1, M8, M9; lower right).
Different colors and types are used to distinguish between the three models.
For example,
blue solid for M2, red dotted fro M1, and green dashed for M3 in the upper left panel.
}
\label{Figure. 6}
\end{figure*}

Since SCs are represented by point-mass particles in the present simulation,
the merging of the SCs with the main cluster
and the subsequent destruction of them can not be investigated
in detail. It is, however, very likely that most of the SCs are tidally destroyed
to form diffuse stellar halo around the cluster 
because SCs follow the following mass-size relation (e.g., Zepf et al. 1999; Larsen 2004):
\begin{equation}
r_{\rm sc} = C_0 m_{\rm sc}^{0.1},
\end{equation}
where $C_0$ is a constant. This means that the mass density of stars
is lower for clusters with lower masses as follows:
\begin{equation}
\rho_{\rm sc}  \propto m_{\rm sc}^{0.7}.
\end{equation}
Therefore, low-mass clusters in a SCC can be completely destroyed by the main cluster
so that their stars can form diffuse stellar halos around the main cluster.
These results are discussed in \S 5.3 in the context of the observed outer halos of GCs.

If all of the low-mass clusters within 5pc from the main cluster at $T=0.28$ Gyr
become the stellar halo in the fiducial model, 
then the mass fraction of the halo to the main cluster
is $\sim 0.06$, which is much larger than the observed mass
fractions (e.g., 0.001 for NGC 1851) of stellar halos in GCs (See Bekki \& Yong 2012 for discussion
on this issue).
Therefore, the present study predicts that (i) stellar halos of the Galactic GCs
were initially denser and more massive and (ii) they have lost
the vast majority of their masses owing to tidal stripping by
the Galaxy  by now. It might be a formidable task for future observations
to detect such a denser and more massive stellar halo around a GC at higher
redshifts. If young massive clusters in nearby galaxies have such
stellar halos,  then they might have been formed as a results of
cluster merging in their host cluster complexes.

The maximum possible mass of the accreted AGB ejecta 
(both from the main cluster and other low-mass ones) within 0.28 Gyr
is $1.3 \times 10^5 {\rm M}_{\odot}$ in the fiducial model.
This is already similar to the present-day typical mass
of SG stars in the Galactic GCs with Na-O anti-correlations (C09).
In the present scenario,  $M_{\rm acc}$ of
{\it  AGB ejecta and cold ISM } can become
as large as $[2-3] \times 10^5 {\rm M}_{\odot}$ for a reasonable set of model parameters,
as described in the following section (\S 4.2).
The  main cluster can lose a significant fraction of its stellar  mass
through (i) mass loss in AGB phases 
and (ii) stripping of stars by the tidal field of its host galaxy
or other luminous galaxies (e.g., the Galaxy).  The final (i.e., present-day) mass of the main cluster 
($m_{\rm mc, f}$) is therefore as follows:
\begin{equation}
m_{\rm mc, f}=(1-f_{\rm strip})(1-f_{\rm ej})m_{\rm mc},
\end{equation}
where $f_{\rm strip}$ is the mass fraction of stars stripped
from the main cluster and $f_{\rm ej}$ is the fraction of stellar mass
that is lost through stellar winds in AGB phases. 
Since $m_{\rm mc}$ is defined as the total mass of the main cluster
when high-mass AGB stars start to eject gas (i.e., not initial
cluster mass before the loss of massive stars through SN explosions),
$f_{\rm ej}$ is $\sim 0.4$ for $\alpha=2.35$
and $\sim 0.6$ for $\alpha=2.05$ (i.e., a top-heavy IMF).
An appropriate value of $f_{\rm strip}$ is 0.4-0.5
for $m_{\rm sc}= 5 \times 10^5 {\rm M}_{\odot}$ at $R=5$ kpc
from the center of the Galaxy.
Thus, if we adopt
$f_{\rm strip}=0.5$ 
and 
$f_{\rm ej}=0.4$,
then  
\begin{equation}
m_{\rm mc, f} \approx 0.3m_{\rm mc}.
\end{equation}
It should be noted here that since gaseous winds
from all AGB stars with different masses are assumed to be ejected from SCs
in this estimation of $f_{\rm ej}$,
this $f_{\rm ej}$ 
is different from that in the equation (1).

This means that if almost all of the accreted
gas can be converted into new stars (i.e., SG stars)
and if the SG stars are not tidally stripped,
then the mass fraction of SG stars can be significant
depending on the IMF of the SG stars. 
For example, the present-day mass of a main cluster 
(FG stars) with $m_{\rm mc}=5 \times 10^5 {\rm M}_{\odot}$ 
is $m_{\rm mc, f}=1.5 \times 10^5 {\rm M}_{\odot}$ 
whereas the total mass of SG stars formed
from the accreted gas  can be as large as
is $2.0 \times 10^5 {\rm M}_{\odot}$ for $m_{\rm mc, f}=3.0 \times 10^5 {\rm M}_{\odot}$,
if only low-mass stars are formed owing to a bottom-heavy IMF.
This means that the  mass budget problem is much less severe
in this scenario. Since this discussion is a bit qualitative,
we will need to investigate the mass budget problem in this scenario
more qualitatively  using a model with different IMFs for FG and 
SG stars and a reasonable range of $f_{\rm strip}$
in our future papers.

\begin{figure*}
\psfig{file=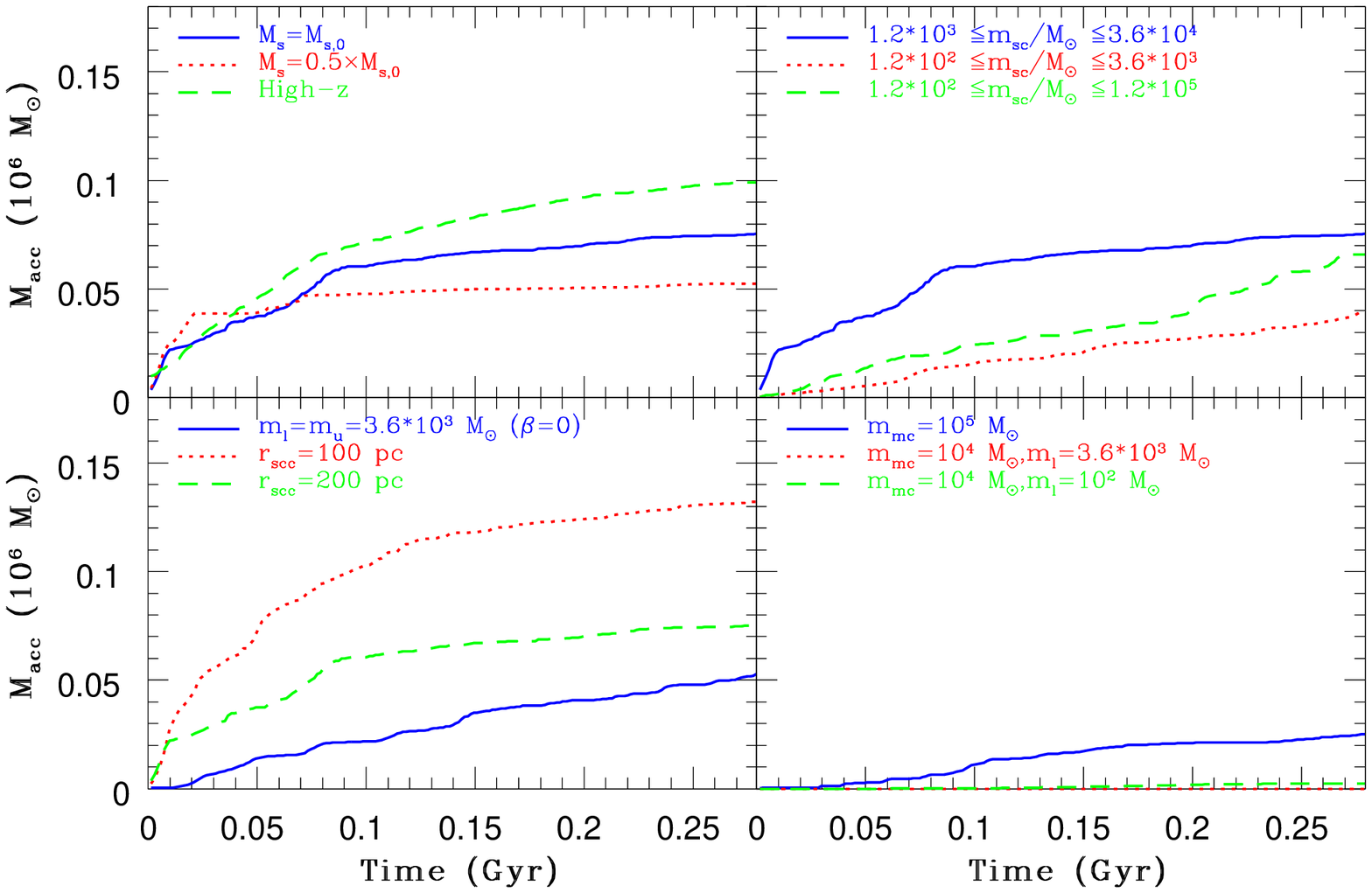,width=16.cm}
\caption{
The same as  Fig. 6 but for  different dwarf models (M1, M10, M11; upper left),
cluster mass ranges in cluster mass functions (M1, M12, M13; upper right),
$\beta$ and $r_{\rm scc}$ (M1, M14, M15; lower left),
and $m_{\rm mc}$ and cluster mass ranges (M16, M17, M18; lower right).
}
\label{Figure. 7}
\end{figure*}

\subsubsection{Parameter dependence}

Fig. 6 shows the following four dependences of $M_{\rm acc}$ on $m_{\rm mc}$
(main cluster's mass),
$R_{\rm scc}$ (SCC's initial position), $r_{\rm mc}$ (main cluster's size),
and $f_{\rm agb}$ (AGB ejecta mass fraction).
First,  more massive main clusters can accrete a larger amount
of AGB eject (i.e., larger  $M_{\rm acc}$), though the ratios of $M_{\rm acc}$ to $m_{\rm mc}$
are not so different between the three models. This $m_{\rm mc}$-dependent result
is expected, because more massive main clusters have deeper gravitational potentials
by which gas can be more efficiently trapped.
Second, there is no significant difference in $M_{\rm acc}$ between
the three models with different $R_{\rm scc}$. In the earlier phases of SCC evolution,
AGB ejecta is more efficiently stripped from the SCC in the model with smaller
$R_{\rm scc}$.
However, $M_{\rm acc}$ can finally become the largest in  the model in which the SCC is born in the
nuclear region of the host dwarf.
Third, main clusters with larger sizes are more likely to accrete more gas from
surrounding SCs. This is mainly because the accretion radius  for 
a main cluster is set
to be the same as the cluster size (i.e., physical condition for gas accretion is 
less strict in the models with large $r_{\rm mc}$).
Fourth, the main cluster
in the model with larger $f_{\rm agb}$ (=0.1) expected from a more top-heavy IMF
can accrete more gas from other SCs. This implies that the IMF of FG stars can
determine the total mass of SG stars.

The mass distributions of SCC-host dwarfs, the lower and upper mass cut-offs
of cluster mass function ($m_{\rm l}$ and $m_{\rm u}$, respectively),
the power-law slope of the cluster mass function ($\beta$), SCC sizes ($r_{\rm scc}$)
can determine $M_{\rm acc}$. Fig. 7 illustrates how $M_{\rm acc}$
depends on these parameters.
First, the high-z dwarf model with a more compact disk shows larger $M_{\rm acc}$
whereas the model with a lower stellar density of the dwarf disk shows
lower $M_{\rm acc}$. 
These are probably
because SCCs are more strongly bound with the deeper gravitational potential wells
so that they can capture more gas from other low-mass SCs.
These results imply that the mass densities 
of SCC-host dwarfs can determine $M_{\rm acc}$ of AGB ejecta in main clusters of SCCs.
Second, the model with a larger number of very low-mass clusters 
(i.e., $m_{\rm l}=1.2 \times 10^2 {\rm M}_{\odot}$) shows lower $M_{\rm acc}$. 
This is firstly because the total mass of the SCs is lower (thus the total mass
of AGB ejecta is lower), and secondly because these clusters can be quickly dispersed
into the field of the host dwarf.
The upper mass cut-off ($m_{\rm u}$) does not influence the final $M_{\rm acc}$
so much.

Third, if all SCs have the same masses of $1.2 \times 10^3 {\rm M}_{\odot}$
(i.e., $\beta=0$), then $M_{\rm acc}$ becomes significantly smaller than
the fiducial model with $\beta=2$. This suggests that hierarhcial cluster distribution
can be important for the gas accretion process. Interestingly,
the model with smaller $r_{\rm scc}$ shows larger $M_{\rm acc}$, which could be
due to the initial more compact distribution of low-mass SCs that can donor gas
to the main cluster.
Fourth, very low-mass clusters with $m_{\rm mc}=10^5 {\rm M}_{\odot}$
and $m_{\rm mc}=10^4 {\rm M}_{\odot}$ 
can have small $M_{\rm acc}$, as expected from their
shallow gravitational potentials.  However, it should be stressed here that
the model with  $m_{\rm mc}=10^4 {\rm M}_{\odot}$ and
$3.6 \times 10 {\rm M}_{\odot} \le m_{\rm sc} \le  1.2 \times 10^3 {\rm M}_{\odot}$,
can accrete AGB ejecta with $M_{\rm acc}=2.5 \times 10^3 {\rm M}_{\odot}$.
On the other hand,
the model with  $m_{\rm mc}=10^4 {\rm M}_{\odot}$ and
$1.2 \times 10^3 {\rm M}_{\odot} \le m_{\rm sc} \le  3.6 \times 10^4 {\rm M}_{\odot}$
can not accrete AGB ejecta at all, because the main cluster is less
massive than a significant fraction of SCs: the main cluster
can not `steal' gas from SCs more massive
than the cluster.
These result imply that even low-mass clusters can have multiple stellar populations,
if they are embedded in clusters of very low-mass clusters.

\begin{figure}
\psfig{file=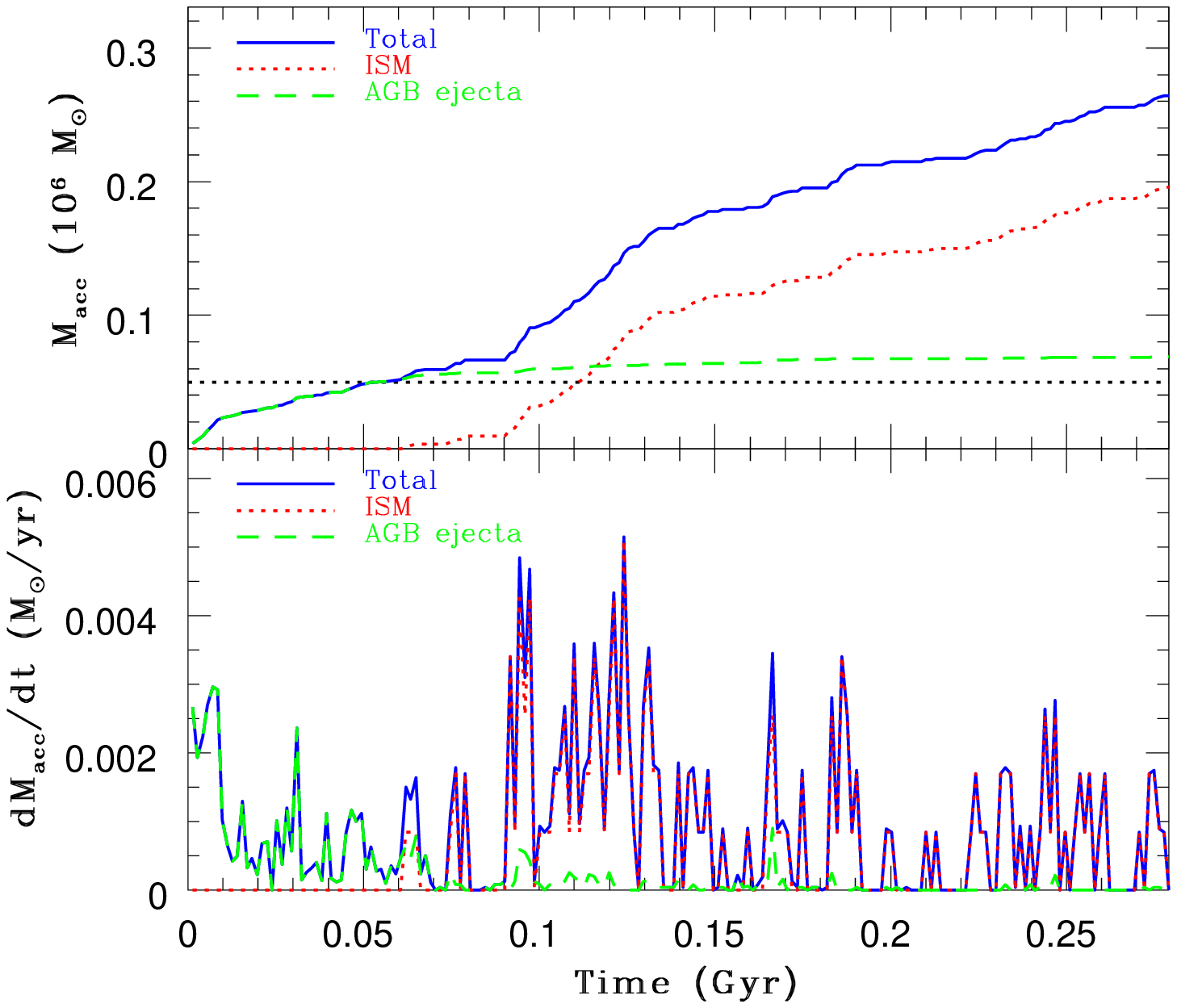,width=8.5cm}
\caption{
The same as Fig.3 but for the model M19 for which the model parameters
are exactly the same as the fiducial model M1 except $M_{\rm g}=0.1$ in this model.
The total gas (blue solid), cold ISM (red dotted),  and AGB ejecta (green dashed)
that are accreted onto the main cluster
are separately shown in this figure.
}
\label{Figure. 8}
\end{figure}

\begin{figure}
\psfig{file=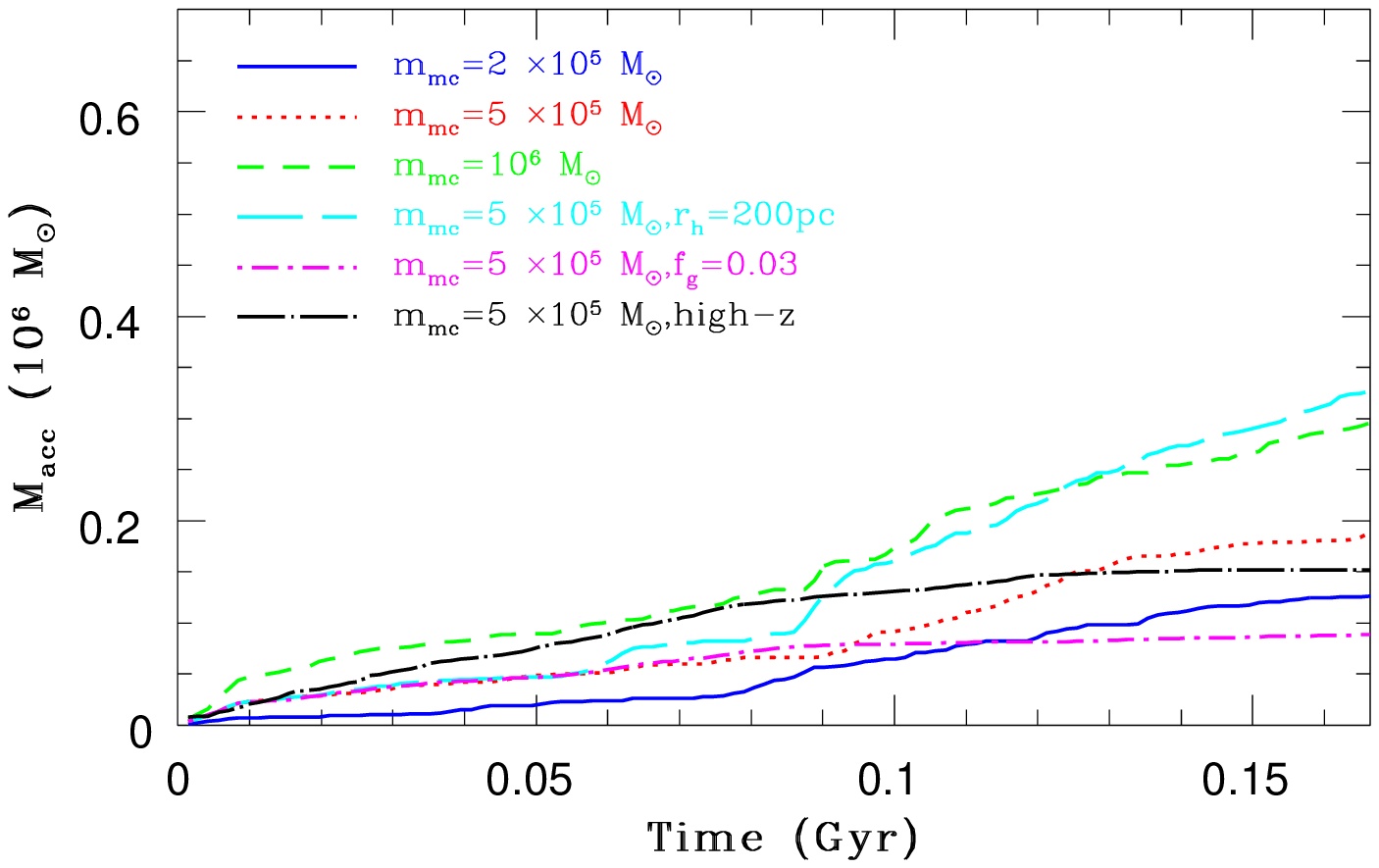,width=8.5cm}
\caption{
Evolution of $M_{\rm acc}$ for different models with cold ISM,
M20 (blue solid), M19 (red dotted),
M21 (green short-dashed), M22 (cyan long-dashed),
M23 (magenta dot-short-dashed), and M24 (dot-long-dashed).
}
\label{Figure. 9}
\end{figure}

\subsection{With cold ISM}

\subsection{Later accretion of cold ISM}

Fig. 8 shows that the gas accretion onto the main cluster is dominated by that of
AGB ejecta in the earlier evolution of the SCC ($T<0.1$ Gyr) in the fiducial model
with cold ISM (M19). Owing to the large gas hole in the initial gas disk of the dwarf
in  this model,  cold ISM, which was escaped from the energetic influence 
of multiple SNe and thus from chemical pollution by the ejecta of SNe,
can start to interact with the SCC later ($T > 0.1$ Gyr). As a result of this,
the rate of gas accretion of cold ISM onto the main cluster  can  dramatically
increase around $T=0.1$ Gyr and shows multiple peaks after $T=0.1$ Gyr. 
The gas accretion rate of cold ISM can become almost always higher than that of AGB ejecta
after $T=0.1$ Gyr, and the total mass of the accreted gas can be 
$2.6 \times 10^5 {\rm M}_{\odot}$ at $T=0.28$ Gyr, which is by a factor of five
larger than the maximum possible $M_{\rm acc}$ expected from self-accretion of
the main cluster itself.

If star formation is possible from the accreted gas in the early evolution of this SCC
($T<0.1$ Gyr), then new (SG) stars can have chemical abundances determined by chemical
yields of AGB ejecta, because no dilution of the ejecta by ISM  is possible.
Accordingly,  the new stars formed earlier can have Na-enhanced 
chemical abundance patters. For example, if the new SG stars are formed from
ejecta of AGB stars with $m_{\rm s}=5 {\rm M}_{\odot}$ and $Z=3 \times 10^{-4}$,
then [Na/Fe]=+0.2 and [O/Fe]=$-0.5$ are expected for the SG stars
for the AGB yield table provided by
Ventura et al. (2013). 
However, Na-enhanced stars do not always show O-depletion,
if they are formed from AGB ejecta (See Fig. 6 in Ventura et al. 2013).
In the later evolution phase of this SCC, the total mass of the accreted ISM
becomes significantly larger than that of the accreted AGB eject. Therefore,
the chemical abundances of new stars formed later can be similar to those of ISM
and FG stars (i.e., existing stars of the main cluster):
these new stars should  have Na-normal  abundance patterns. Thus,
the older SG stars have higher [Na/Fe] than the younger 
SG stars in this scenario (surely older and younger SG stars are younger
than FG stars).
Chemical abundances of cold ISM initially outside the gas hole of the SCC
are assumed to be the same as those of FG stars in the main cluster in this scenario.
However,  it is possible that the chemical abundances of the ISM could have changed
significantly over 0.2 Gyr (i.e., time lag between the formation of the SCC
and the commencement of ISM accretion). If the metallicity (e.g., [Fe/H]) evolution due to 
star formation in ISM over 0.2 Gyr is well less than 0.05 dex, then this scenario
would be still viable. This point will needs to be addressed in our future papers.

\subsection{Parameter dependence}

Fig. 9 demonstrates that $M_{\rm acc}$ in the models M19-M24
can be significantly different depending
on $m_{\rm mc}$, $r_{\rm h}$, $f_{\rm g}$, and the stellar distributions of dwarfs:
$M_{\rm acc}$ depends on these parameters as follows.
First, $M_{\rm acc}$ can be larger in the model with larger $m_{\rm mc}$, as the
models without cold ISM, though the ratio of $M_{\rm acc}$ to $m_{\rm mc}$ 
is not so different between the three models with different $m_{\rm mc}$ (M19, 20, and 21).
It should be noted here that $M_{\rm acc}$ can be $1.5 \times 10^5 {\rm M}_{\odot}$
(corresponding to the total stellar mass of SG stars in a typical Galactic GC)
in the main cluster of the model with 
$m_{\rm mc}=2 \times 10^5 {\rm M}_{\odot}$.
Second, the model with a smaller gas hole ($r_{\rm h}=200$ pc) shows 
large $M_{\rm acc}$ and earlier commencement of ISM accretion onto the main cluster.
The density of ISM around the main cluster can become higher from the early evolution
phase of the SCC in this model so that the Bondi-type gas accretion
(Bondi 1952) in the main cluster
can become quite efficient in this model.

Third, $M_{\rm acc}$ is lower in
the model with a lower gas fraction (thus lower gas density) of the dwarf,
because the efficiency of the Bondi-type gas accretion depends on the density
of gas surrounding the accreting object.
This result suggests that the gas mass fraction of a gas-rich dwarf galaxy
can determine the mass fraction of SG stars in a GC formed in the galaxy.
Fourth, the main cluster in the high-z more compact dwarf model can have 
$M_{\rm acc}$ larger than that in the fiducial (less compact) model with cold ISM.
It is confirmed that the models M25, M26, and M27, in which dynamical friction
timescale is longer owing to the lower stellar mass densities of SCC-hosting 
dwarfs, show similar amount of $M_{\rm acc}$: M27 model shows 
$M_{\rm acc} = 2\times 10^5 {\rm M}_{\odot}$.

The large $M_{\rm acc}$ ($ \sim [0.2-0.3] \times m_{\rm mc}$)
derived in some models with ISM implies that
the mass budget problem is much less serious  in the GC formation scenario from SCCs.
However, the scenario has not yet provided a complete solution to the problem,
because the physical origins for the required high star formation efficiency in SG 
star formation 
and preferential formation of low-mass SG stars are not clear in the scenario.
Furthermore, it remains unclear 
whether $M_{\rm acc, agb}$, $M_{\rm acc, ism}$,
and the time evolution of these derived in the present simulations
can explain the observed distributions of [Na/Fe] and [O/Fe] in the Galactic GCs
with multiple stellar populations, if SG stars can form from the mixed gas
of AGB ejecta and ISM.
These will need to be investigated extensively in our future studies.

\section{Discussion}

\subsection{The mass budget problem}

In the present new GC formation scenario from SCCs,
the  main cluster within a SCC
can accrete more than $10^5 {\rm M}_{\odot}$ gas from  AGB stars of the SCC,
if $M_{\rm scc}$ is as large as $10^7 {\rm M}_{\odot}$.
This means that if SG stars can be formed from the gas with a very
high star formation efficiency,
the mass budget problem can be much less severe in the scenario.
However, the timescale of gas accretion onto forming GCs 
is quite long ($\sim 10^8$ yr), which is longer than the lifetimes
of massive stars that explode as SNe. Therefore,
it is possible that SNe from SG stars can truncate gas accretion
onto forming GCs owing to their energetic feedback effects. 
This possible truncation of SG star formation by earlier 
SNe of SG stars themselves was already pointed by D'Ercole et al. (2010).
Thus,
the mass budget problem can be  less severe in the new scenario only
if the IMF of SG stars is top-light: the upper mass cut-off of the IMF
should be quite low so as to suppress the formation of massive SNe.
Our forthcoming  papers based on the GC formation scenario will discuss
this IMF issue in detail.

The new SCC scenario does not need to assume very massive 
single star clusters ($m_{\rm sc} \ge 4 \times 10^6 {\rm M}_{\odot}$)
as progenitor of GCs. Accordingly, 
efficient stripping  of almost all (more than 90\%) FG stars
by GC host galaxies 
is not required either to explain the observed large fraction (70\%)
of SG stars in the Galactic GCs.
FG stars can be more efficiently stripped by GC host galaxies
than SG stars in the new scenario too,
but the difference in the mass fractions of stripped stars
between FG and SG stars is much smaller in the new scenario: only a factor of two
decrease in mass is required to explain the  present-day typical GC mass and 
the fraction of SG stars. Such a moderate mass decrease is consistent with
theoretical predictions on the fraction of
GC stars stripped by the Galaxy (e.g., Vesperini 1997; Rossi et al. 2016).
The new scenario, however,  still assumes that the initial total mass
of a SCC is  large ($M_{\rm scc}=10^6-10^7 {\rm M}_{\odot}$)
owing to its formation within a massive GMC or a GMC complex.
Therefore, it appears to have a problem in explaining the observed
small fraction of field stars with metallicities similar to those of GCs
in the Fornax dwarf galaxy (Larsen et al. 2012).

The Fornax dwarf galaxy might have interacted violently with
the Galaxy and even merged with other dwarf galaxies in the
Galactic halo (e.g., Coleman et al. 2005; Yozin \& Bekki 2012;
del Pino et al. 2015). Therefore,
it might have lost a large fraction of its initial dark matter 
and  field stars through
the past interaction and merging.
Unlike field stars, GCs can sink into the central region of
its host dwarf 
through dynamical friction against the host's field stars
so that they can become 
closer to the host's center.
It is thus possible that field stars of the Fornax dwarf
have been preferentially lost to the Galactic  halo through
tidal stripping.
Even if the Fornax dwarf initially had FG stars stripped from their GCs,
such FG stars could have already stripped from the dwarf
to form the Galactic halo field
stars. Thus, the observed small fraction of low-metallicity 
field stars in the Fornax dwarf (Larsen et al. 2012) would not
be a serious problem for the new scenario.
Since the above discussion is only qualitative and slightly speculative,
we will need to investigate the evolution
of the mass fraction of low-metallicity field stars from original
GCs in the Fornax for the realistic 3D orbits of the Fornax around the Galaxy.

Niederhofer et al. (2016) have recently discovered chemical abundance spreads
in intermediate-age GCs (Lindsay 1, NGC 416, and NGC 339) with ages
ranging from 6 Gyr to 7.5 Gyr in the Magellanic Clouds. This discovery
strongly suggests that multiple stellar populations are not limited to
old ($>10$ Gyr) Galactic GCs, though the mass fractions of SG stars 
($<0.45$) are lower in these GCs than in the Galactic GCs. 
The observed mass fractions however required large original masses
of these GCs in any self-enrichment scenario. 
It is not clear, however, 
whether 
massive SCCs  ($M_{\rm ssc}=10^6-10^7 {\rm M}_{\odot}$)
can be formed within the Magellanic clouds about 7 Gyr in the present
scenario. 
It is our future work to investigate how and why such intermediate-age clusters
with multiple stellar populations could be formed within
the Magellanic Clouds based on our improved SCC model.

\subsection{Dilution of AGB eject by cold ISM initially outside giant gaseous holes}

Time evolution of gas accretion from AGB stars and cold pristine gas
are freely chosen so that the observed anti-correlations
between light elements in GCs can be reproduced in previous chemical evolution models
of GC formation (e.g., Bekki et al. 2007; D'Ercole et al. 2010).
However, 
such evolution of gas accretion rates 
can not be so freely changed
in the  new scenario.
SCCs are still embedded in giant gaseous holes 
when AGB stars start to eject gas through their stellar winds
in the scenario.
Accordingly,  it takes longer time for cold gas initially outside the holes
to be accreted onto the main clusters.
The time lag between accretion of AGB ejecta and that of cold gas
can be longer in more massive GCs, because bigger gaseous holes
can form in more massive GCs with a larger number of energetic massive stars and
SNe. 
The new scenario therefore provides the following predictions on the
chemical abundances of GCs with multiple stellar populations.
First,  Na-enhanced stars  are slightly older than
Na-normal stars in GCs,
because AGB ejecta with Na-enhanced  abundances 
can be accreted onto the main clusters earlier than cold gas
and then be converted into new stars. Na-normal
stars can form only after a significant amount of cold gas is accreted onto
the main clusters to dilute the existing AGB ejecta.

Second, a significantly larger amount of
pure AGB ejecta accreted onto  main clusters  can be
converted into  star formation before the accretion
of pristine cold gas 
in massive GCs, because accretion of cold gas on main clusters
is much delayed in SCCs containing more massive main clusters
(i.e., due to larger gaseous holes).
Stars formed from AGB ejecta that is not
mixed with pristine cold gas can have higher helium (He)  abundances ($Y$),
because AGB ejecta can have such $Y$ (e.g., Ventura \& D'Antona 2005).
Therefore the new scenario can  explain the presence of subpopulations 
with  large He abundances observed
in NGC 2808 and omega Cen.
Third, although low-mass GCs formed from low-mass SCCs can accrete only a small amount
of AGB ejecta,  they can still accrete some amount of
cold gas from ISM in gas-rich dwarfs to form new stars within them. 
These low-mass GCs can therefore contain SG stars with less enhanced N.
These predictions are based only the present hydrodynamical simulations of 
gas accretion in GCs, and accordingly will need to be confirmed in our future 
more sophisticated numerical simulations of GC formation with chemical evolution.

It should be noted that ISM diluting AGB ejecta is assumed to have
metallicities almost the same as those of FG stars in the above discussion.
Since the present study did not investigate the chemical evolution of ISM of 
a GC-hosting dwarf, it can not predict the chemical abundances of ISM around
forming GCs.  It could be possible that cold ISM around a forming GC
can have different metallicities owing to active formation of field stars
around and within the ISM
during the GC formation.
It is our future study to investigate  whether chemical
abundances of ISM around forming GCs 
can be quite similar to those of FG stars of the GCs.
If their chemical abundances are different from those of FG stars
by more than 0.05 dex, then we
will either  discard the present GC formation scenario
or need to consider other sources that can dilute AGB ejecta.

\begin{figure}
\psfig{file=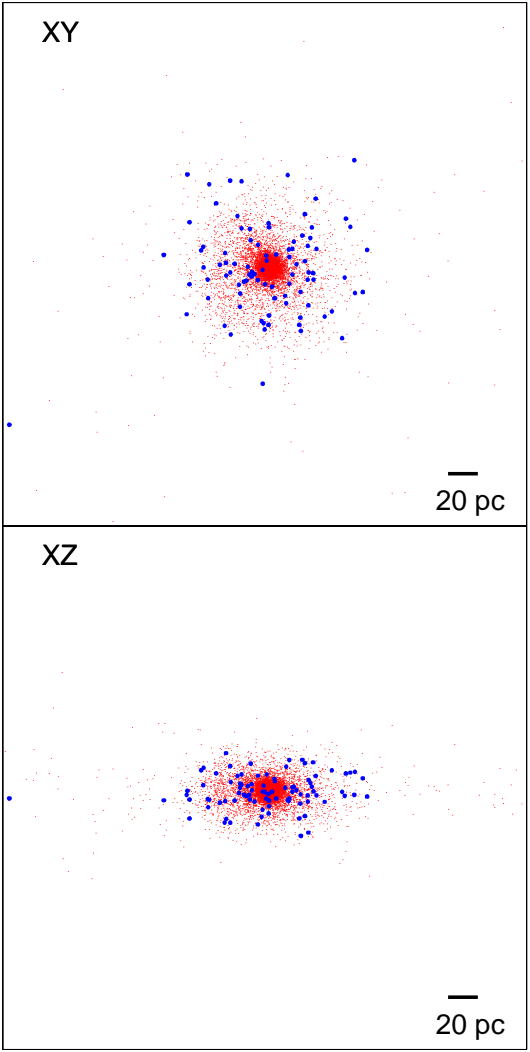,width=8.5cm}
\caption{
Distributions of stars in the main cluster (small red dots)
and new stars formed from AGB ejecta of low-mass clusters 
(big blue dots) projected onto
the $x$-$y$ (upper) and $x$-$z$ planes (lower). This model is exactly the same
as the fiducial model M1 except that field AGB stars are not included
just for clarity.
}
\label{Figure. 10}
\end{figure}

\subsection{Stellar halos around GCs from destruction of low-mass clusters}

Recent observations discovered diffuse stellar halos around a number of GCs,
such as NGC 1851
(Olszewski et al. 2010)
in Whiting 1 (Carraro et al. 2007),
AM 4 (Carraro et al. 2009),
NGC 5694
(Correnti et al. 2011),
and NGC 1904 (Carballo-Bello \& Mart\'inez-Delgado 2011),
and M2 (Kuzma et al. 2016).
The stellar halos around NGC 1851 and M2 are rather large (200-500pc) with
spherical shapes and power-law density distributions,
thought the mass fractions of them are quite minor (0.1\% and 1.6\%, respectively).
It is not so clear whether only a fraction of GCs in the Galaxy can host such
outer diffuse components or not.
The physical process that forms the stellar halos is not understood yet.
One of possible scenarios for the formation of the halos is that these GCs with
halos were formed from nuclei of nucleated dwarfs that were destroyed
by the Galaxy many Gyr ago (e.g., Bekki \& Yong 2012;
Bekki \& Tsujimoto 2016; Kuzma et al. 2016). The stripped nuclei (=GCs) can still 
have old field stars that initially constituted the host dwarf galaxies.
There could be some differences in chemical abundances between GCs and their stellar
halos, because GCs and field stars could have formed in different areas of their host
dwarfs.

The present scenario offers an alternative explanation for the origin of
such stellar halos around GCs. A massive star cluster that is 
the progenitor of a GC  should be a part of 
clustering of star clusters in hierarchical star formation: such a massive cluster
should not be an isolated single cluster.
During the dynamical evolution of the SCC,
some low-mass clusters are stripped from the surrounding of the main massive cluster
whereas some merge with the main cluster and consequently
are destroyed by the cluster.  Stars initially in the destroyed clusters can constitute
the diffuse stellar halo around the main cluster. In this scenario,
chemical abundances of the halo is almost exactly the same as those of the main cluster,
which is in a striking contrast with the above stripped nuclei scenario.
Thus, future observations on the chemical abundances of stars both for
halos and main components of GCs
will be able to discriminate between these two scenarios.

\subsection{Secondary star formation ?}

It is our future study to investigate the formation of new stars
(SG stars)  from gas accreted
onto the main clusters of SCCs for various model  in detail,
because the star formation model adopted in the present study
is appropriate only for galaxy-wide star formation (not for SG formation
within dense clusters).
However, it would be instructive for the present study
to describe briefly the time evolution of SFR for SG stars
and the spatial distribution of the SG stars.
Fig. 10 shows that the distribution of FG and SG stars in the main cluster
of a SCC in a model in which the model parameters are exactly the same as
the fiducial model M1 except (i) no gas ejection of field AGB stars and (ii)
no SN feedback effects. Therefore, the SG stars are those formed from
AGB ejecta of low-mass SCs only and SN explosion  can not influence the SF
history. In this comparative model,  $M_{\rm acc}$ is slightly
larger than that derived in the fiducial model.
The mass of SG star particle ranges from $5.3 \times 10^3 {\rm M}_{\odot}$
to $1.7 \times 10^4 {\rm M}_{\odot}$, and the total mass of the SG stars
is $8.6 \times 10^4 {\rm M}_{\odot}$.

Clearly, the distribution of SG stars does not show a strong central concentration,
which is significantly different from our previous results  (B11) on the more
compact distribution of SG stars in a single massive cluster.
This result implies that SG stars do not necessarily have more compact distributions
than FG stars in the new scenario.
The star formation rate for SG stars is kept low 
($\sim 1 \times 10^{-4} {\rm M}_{\odot}$ yr$^{-1}$) for $T<0.1$ Gyr,
and it has a peak of $0.001 {\rm M}_{\odot}$ yr$^{-1}$ in a weak bust phase
at $T=0.13$ Gyr. These lower SFR might be difficult to be detected observationally
in massive young star clusters with secondary star formation. 
The spatial resolution of the present simulation is  0.47pc at most, which
implies that the above results could be due partly to the simulations resolution
that might not allow this study to investigate the subpc-scale SF physics in detail.
We will address this important issue on the spatial distributions of SG stars
in GCs in our future papers.

\begin{table*}
\centering
\begin{minipage}{160mm}
\caption{Description of the key physical processes
of GC formation in the present GC formation
scenario from star cluster complexes (SCCs).}
\begin{tabular}{lll}
{ Physical process
 } & 
{ Papers  \footnote{ If the listed physical process is investigated
in our previous paper or in the present paper, then the reference of the paper
is shown. If not, the mark `-' is shown, which means that the 
physical process will be investigated in our forthcoming papers.
 }} & 
{ Comments \footnote{  The potential serious problems of the new scenario
are listed if necessary.
}}  \\
Formation of massive gas clumps  in  dwarfs  & B15a, B16  &  \\
Fractal structures in gas clumps   &  $-$   &  \\
Formation of SCCs within  gas clumps   &  $-$   &  Survival of hierarchical 
structure for more than $10^7$ yr is crucial.  \\
Merging of SCs within SCCs   & This work   &  Future models
need to investigate disintegration of small SCs.  \\
Ejection of AGB ejecta    & This work   &  \\
Accretion of AGB ejecta onto forming GCs    & This work   &  \\
Accretion  of pristine ISM    & This work   &  \\
Chemical mixing  of AGB ejecta and ISM    &  $-$   &  Dilution of AGB ejecta by ISM needs to be modeled self-consistently.\\
Secondary star formation   & $-$   &  SNe should be suppressed in SG star
formation. \\
Chemical enrichment of ISM by field stars  & $-$  &  Small metallicity difference
between ISM and AGB ejecta is required. \\
Formation  of giant HI holes by SNe  & $-$  &  \\
\end{tabular}
\end{minipage}
\end{table*}

\section{Conclusion}

We have proposed a new GC formation scenario in which star cluster complexes
(SCCs) can form GCs with multiple stellar populations in gas-rich dwarf galaxies.
In the scenario,  a present-day GC was initially the most massive star cluster
(`main cluster')  embedded in a loosely
bound star cluster complex (SCC) consisting of numerous stellar associations and low-mass
star clusters (SCs) that were formed almost simultaneously with the main cluster.
The SCC resides in the central region of a giant gas hole (i.e, local region
devoid of gas) that was created by energetic massive stars and SNe of the SCC itself.
Gas ejected from AGB stars in low-mass SCs and stellar
 association of the SCC
is accreted onto the main cluster and then converted into new stars (i.e., the second
generation of stars; SG) in the scenario. 
The physical processes that have been/will be  investigated in the present 
or future studies are briefly summarized with some comments on
the possible serious problems of the new scenario.

A crucial question in the scenario is whether
an enough amount of AGB ejecta 
($\sim 10^5 {\rm M}_{\odot}$) can be accreted onto the main cluster from 
individual low-mass SCs and stellar associations before
they are tidally stripped from the main cluster
and destroyed/disintegrated to become field stars in the GC host dwarf galaxy.
We have investigated this question using hydrodynamical simulations
of the evolution of AGB ejecta in gas-rich dwarfs for a wide range of model parameters.
Key parameters are the initial mass of a main cluster ($m_{\rm mc}$),
the distance of the SCC from the center of its host dwarf galaxy ($R_{\rm scc}$),
the fraction of AGB ejecta in a star cluster ($f_{\rm agb}$),
the total mass of gas initially in the dwarf disk ($M_{\rm g}$),
and structure parameters of the dwarf.
The principal results are as follows. \\

(1) Main clusters can accrete a significant amount of AGB ejecta from other low-mass clusters 
in SCCs
before most of the low-mass clusters are stripped from the surroundings of the main clusters
in GC-host dwarfs.  The total mass of the accreted AGB ejecta ($M_{\rm acc}$) 
can be as large as 
$\sim 10^5 {\rm M}_{\odot}$ for $m_{\rm mc} = 5 \times 10^5 {\rm M}_{\odot}$.
This `donation' of AGB ejecta from other member clusters in a SCC is significant,
because there are numerous low-mass SCs in a SCC owing to the power-law cluster mass function
with a power-law index of $\beta=2$. 
This result suggests  that hierarchical star formation, which is a main physical origin
for the power-law cluster mass function,  needs to be considered in discussing
the origin of GCs with multiple stellar populations.  \\

(2) The total masses of AGB ejecta accreted onto main clusters 
are  larger in  the models with larger  $m_{\rm mc}$ for a given set of 
other model parameters, though
the ratio of $M_{\rm acc}$ to  $m_{\rm mc}$ is not so different
for $m_{\rm mc} > 10^5 {\rm M}_{\odot}$.  Low-mass main clusters with
$m_{\rm mc} < 10^5 {\rm M}_{\odot}$ can not accrete AGB ejecta so efficiently from
other clusters as massive main clusters for a given cluster mass function.
These results imply that $m_{\rm mc}$ is a key
parameter that controls the total mass of SG stars which can  form from gas
accreted onto the main clusters.   \\

(3) The IMF-dependent parameter  $f_{\rm agb}$ can determine the time evolution
of the accretion rates of AGB ejecta and thus the total amount of the accreted gas
($M_{\rm acc}$).
A larger amount of AGB ejecta can be accreted onto main clusters in the models
with larger $f_{\rm agb}$, which means that $M_{\rm acc}$
is larger for SCCs with more top-heavy IMFs. 
The locations of SCCs within their host dwarfs can also influence the 
the accretion processes of AGB ejecta. 
Main clusters with smaller sizes can accrete a less amount of AGB ejecta,
mainly because the accretion radius is smaller. \\

(4) Stellar structures of SCC-host dwarfs, cluster mass functions (e.g., $\beta$,
$m_{\rm l}$, and $m_{\rm u}$),  and SCC sizes can determine $M_{\rm acc}$ of the main
clusters in SCCs. The model with $\beta=0$ (no hierarchy in SCC) shows significantly
smaller $M_{\rm acc}$ than the fiducial model with $\beta=2$, which implies 
that hierarchical star formation, which determines  $\beta$,  is important
for the gas accretion process. Low-mass main
clusters with $m_{\rm mc}=10^4 {\rm M}_{\odot}$ can accrete AGB ejecta from
other low-mass clusters, if $m_{\rm l} \sim 10^2 {\rm M}_{\odot}$
and $m_{\rm u} \sim 3\times  10^3 {\rm M}_{\odot}$. 
  \\

(5) Cold gas initially outside large gas holes can be accreted onto the main clusters
of SCCs later than AGB ejecta. This time lag between the accretion of AGB ejecta
and that of cold ISM suggests that SG stars formed earlier in main clusters
can show Na-enhanced  abundance patterns.
This results also implies that stars with Na-enhanced 
sub-populations in GCs should be younger than those with Na-normal 
stars. 
If AGB ejecta is  helium-rich ($Y \sim  0.35$) and if secondary star formation
occurs within 0.1 Gyr after the commencement of gas accretion of AGB ejecta onto main clusters
(i.e., well before the accretion of cold ISM on the clusters),
then the clusters can have sub-populations with rather high Y.
If the host dwarf galaxies are gas-rich, then the total amount of AGB ejecta
and cold gas accreted
onto the main clusters can be as large as $[2-3] \times 10^5 {\rm M}_{\odot}$ 
for $m_{\rm mc}=5 \times 10^5 {\rm M}_{\odot}$. \\

(6) The total amount of AGB ejecta and cold ISM accreted onto the main clusters
in the models with ISM depends on model parameters in a similar way as described 
for the models without ISM. The total mass of accreted ISM ($M_{\rm acc, ism}$)
is appreciably larger
than that of accreted AGB ejecta ($M_{\rm acc, agb}$) for most model in the present study. 
It is not clear whether the derived mass ratio of $M_{\rm acc, ism}$ to $M_{\rm acc, agb}$
is consistent with the chemical abundance distributions of light elements observed
in the Galactic GCs with multiple stellar populations. This point will be investigated
in our future studies. \\

(7) Low-mass SCs can merge with the main clusters in SCCs. Since
the mass densities of the SCs is much lower than the more massive main clusters
owing to the mass-size relation ($r_{\rm sc} \propto m_{\rm sc}^{0.1}$),
they can be destroyed by the main clusters to form diffuse stellar halos. Only
a small fraction of the stars can add their masses to the main clusters.
This result suggests that GCs formed in the SCC scenario might have diffuse stellar halos
just after their formation (i.e., after the dispersal of SCCs). 
This result also implies that the origin of 
the observed stellar halos in some GCs  can be understood in terms of destruction of
low-mass clusters in the GC formation from SCCs.
  \\

(8) The new SCC scenario with external gas accretion has
predictions that are quite different from those 
of previous GC formation scenarios with self-accretion only
(or `self-enrichment' scenario).
For example,
the 'donation' of AGB ejecta from other member clusters in SCCs can potentially
solve the mass
budget problem of GCs with multiple stellar populations.
Also, the scenario naturally predicts
the later accretion of cold ISM onto main clusters  and the formation of stellar
halos around GCs. 
Older SG stars are likely to have higher [Na/Fe] than younger SG stars
in the present scenario, though such an age-[Na/Fe] relation in SG stars depends on the 
adopted AGB yields.

The present study is a very first step toward the better understanding 
of  GC formation from SCCs
with $\beta=2$ that are expected to form in hierarchical star formation.
Since the present study has investigated only the  gas accretion processes in forming
GCs, there are many other issues that remain to be investigated.
In particular, the observed anti-correlations between light elements will need
to be investigated in our future studies with more sophisticated simulations
with chemical evolution in forming GCs.

\section{Acknowledgment}
I (Kenji Bekki; KB) am   grateful to the referee  for  constructive and
useful comments that improved this paper.

\end{document}